\documentclass[a4paper,11pt]{article}
\pdfoutput=1
\usepackage{amsmath,amssymb,xymtex,units}

\usepackage{jinstpub} 

\usepackage{xcolor}
\usepackage{graphicx,epstopdf}
\usepackage{tikz}
\usepackage{lineno} 
\usetikzlibrary{angles, arrows.meta, quotes}
\usepackage{dcolumn}
\usepackage{bm}
\usepackage{ulem}
\usepackage{adjustbox}
\usepackage{multirow}

\abstract{The Multipurpose Detector (MPD) is an experimental array, currently under construction, designed to study the nuclear matter created during the collisions that will be provided by the Nuclotron-based Ion Collider fAcility (NICA) at JINR. The MPD-NICA experiment consists of a typical array of particle detectors as those used to study heavy-ion collisions at LHC and RHIC. To increase the trigger capabilities of MPD for stage 2 of NICA operation, a detector constituted by two arrays of 80 plastic scintillator cells each located symmetrically at opposite sides of the interaction point of MPD is proposed (BeBe). Based on Monte Carlo simulations, a discussion of the potential physics performance of BeBe detector is given for triggering tasks and for the resolution in the determination of the event plane reaction and the centrality of the collisions at NICA. Also, laboratory measurements to estimate the time resolution of individual BeBe cells prototypes are presented. It is shown that a time resolution between 0.47 and 1.39~ns can be reached depending on the number of photosensors employed to collect the scintillation photons. The BeBe detector will be complementary to FFD and FHCAL forward detectors.

} 


\arxivnumber{2110.02506} 

\begin{document}

\title{Performance of BeBe, a proposed dedicated beam-beam monitoring detector for the MPD-NICA experiment at JINR}


\author[a]{Marco Alberto Ayala-Torres}
\author[b,d]{, Lucina Gabriela Espinoza Beltr\'an}
\author[a]{, Marcos Aurelio Fontaine Sanchez}
\author[b]{, Luis A. Hern\'andez-Cruz}
\author[a]{, Luis Manuel Monta\~no}
\author[b]{, Braian Adair Maldonado Luna}
\author[b]{, Eduardo Moreno-Barbosa}
\author[b,f]{, Lucio F. Rebolledo-Herrera}
\author[b]{, Mario Rodr\'iguez-Cahuantzi,$^1$\note{Corresponding author.}}
\author[b,e]{, Valeria Z. Reyna-Ortiz}
\author[b]{, Guillermo Tejeda-Mu\~noz}
\author[b,c]{and C. H. Zepeda Fern\'andez}


\affiliation[a]{Centro de Investigación y de Estudios Avanzados del Instituto Politécnico Nacional (Cinvestav), Mexico City, Mexico}
\affiliation[b]{Facultad de Ciencias Físico Matemáticas, Benemérita Universidad Autónoma de Puebla, Av. San Claudio y 18 Sur, Edif. EMA3-231, Ciudad Universitaria 72570, Puebla, M\'exico}
\affiliation[c]{C\'atedra CONACyT, 03940, CdMx M\'exico}

\affiliation[d]{Facultad de Ciencias F\'isico-Matem\'aticas, Universidad Aut\'onoma de Sinaloa, Avenida de las Am\'ericas y Boulevard C.P. 80000, Culiac\'an, Sinaloa, M\'exico}
\affiliation[e]{Institute of Physics, Jan Kochanowski University, 25-406 Kielce, Poland}
\affiliation[f]{Instituto de Ciencias Nucleares, Universidad Nacional Aut\'onoma de M\'exico, Apartado postal 70-543, CDMX 04510, M\'exico}
\emailAdd{mario.rodriguez@correo.buap.mx}

\maketitle

\flushbottom

\section{Introduction}
\label{sec:introduction}

In collider experiments, the use of a detector capable of monitor the beam activity is desirable. The information provided by particle detectors of this kind is used to identify and discriminate beam-beam minimum bias or centrality events from background and beam-gas interactions. These detectors can be used for the reconstruction of physical observables of interest in heavy-ion collisions such as multiplicity of charged particles, key observable for the determination of the centrality of the collision events and the event plane resolution, and luminosity measurements for determining the absolute cross-section of specific reaction processes. Experiments like PHENIX at RHIC ~\cite{Allen:2003zt} and ALICE at LHC ~\cite{Aamodt:2008zz} have successfully employed particle detectors based on plastic scintillators to generate a minimum bias trigger signal and to monitor the beam activity.

To extend the QCD phase diagram in the richest baryon region, with respect to LHC and RHIC heavy-ion experiments, the Multi-Purpose Detector (MPD-NICA) ~\cite{Golovatyuk:2016zps} is under construction at JINR  where heavy nuclei will collide at $\sqrt{s_{NN}}=4-11$ GeV \cite{Golovatyuk:2019rkb} for Bi+Bi and Au+Au beam species. The planned physics studies of the MPD Collaboration comprise the characterization of the nuclear matter produced in heavy-ion collisions through anisotropic flow measurements, electromagnetic and hard probes as well as the measurement of global observables of the charged particles produced at NICA such as multiplicity and mean transverse momentum, among others. A recent overview of the current status of the physics performance studies of MPD can be found in ~\cite{MPD:2022qhn}.

To perform all of these studies, it is crucial to develop dedicated particle detectors for online beam monitoring and triggering tasks. These types of systems are employed for offline determination of the event reaction plane and collision centrality, two key observables in the study of the nuclear matter produced in heavy-ion collisions. This is the case of the Beam-Beam monitoring detector (BeBe) ~\cite{Alvarado:2018gbb}, a proposed system to increase the trigger capabilities of the MPD detector. It is expected that BeBe contribute to the discrimination of beam-gas interactions from beam-beam collision events and also in the determination of the centrality and reaction plane in heavy-ion collisions at MPD-NICA. The BeBe detector is planned to be installed for stage 2 of NICA operations.

In the following of this work it will be shown that the BeBe detector can be used to generate a trigger signal for MPD with a time resolution between 0.65 and 1.48 ns (sections ~\ref{sec:BeBe_corners} and \ref{sec:Measurements}) for online luminosity measurements of NICA beam (subsection \ref{sec:BeBeTrigger}). In fact, the BeBe trigger efficiency is larger than 95\% for both proton+proton and Bi+Bi/Au+Au collisions at a center of mass energy of 9 and 11 GeV respectively. Moreover, the maximum event plane resolution given by BeBe, for the 1st harmonic, is of the order of 44\% for an impact parameter range from 6 to 11~fm (subsection \ref{sec:BeBecentrality}). The resolution of the centrality determination by BeBe detector of the expected collisions at NICA is 0.05-0.1 for centralities percentages between 20\% and 100\%. Indeed, as previously reported in ~\cite{Kado:2020evi}, BeBe compensates for the low trigger efficiency in low multiplicity proton+proton collision events given by the Fast Forward Detector~(FFD) ~\cite{Yurevich:2013tra}. In this work, it is shown that BeBe also compensates for the decrease of the resolution of the centrality determination for peripheral collisions given by the Forward Hadron Calorimeter~(FHCAL) \cite{Kurepin:2019bmh} (subsection \ref{sec:BeBeResolcentrality}).

\section{BeBe general description}
\label{sec:BeBeDescription}

\begin{figure}[hbt!]
\centering
\includegraphics[scale=0.24]{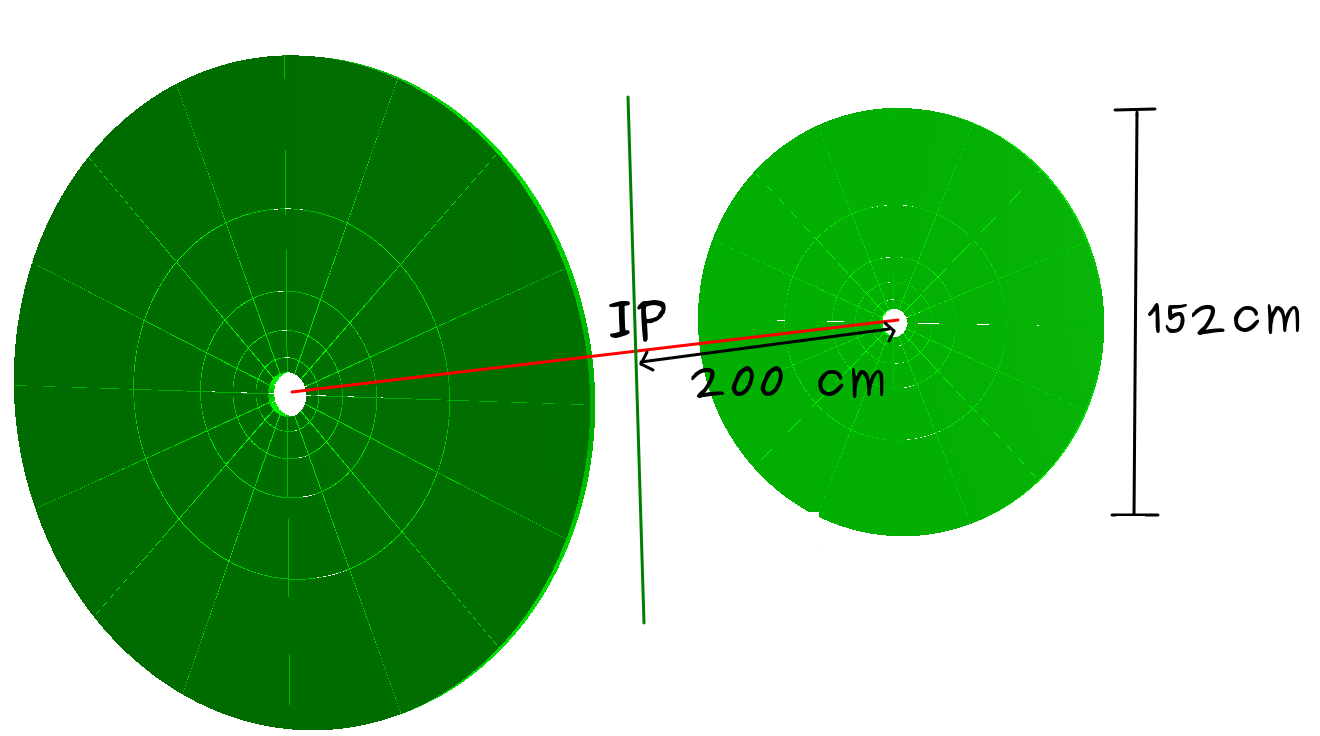}
\caption{BeBe design geometry as rendered by the MPD offline environment.}
\label{fig:BBGeometryIP}
\end{figure}

\begin{figure}[hbt!]
\centering
\includegraphics[scale=0.45]{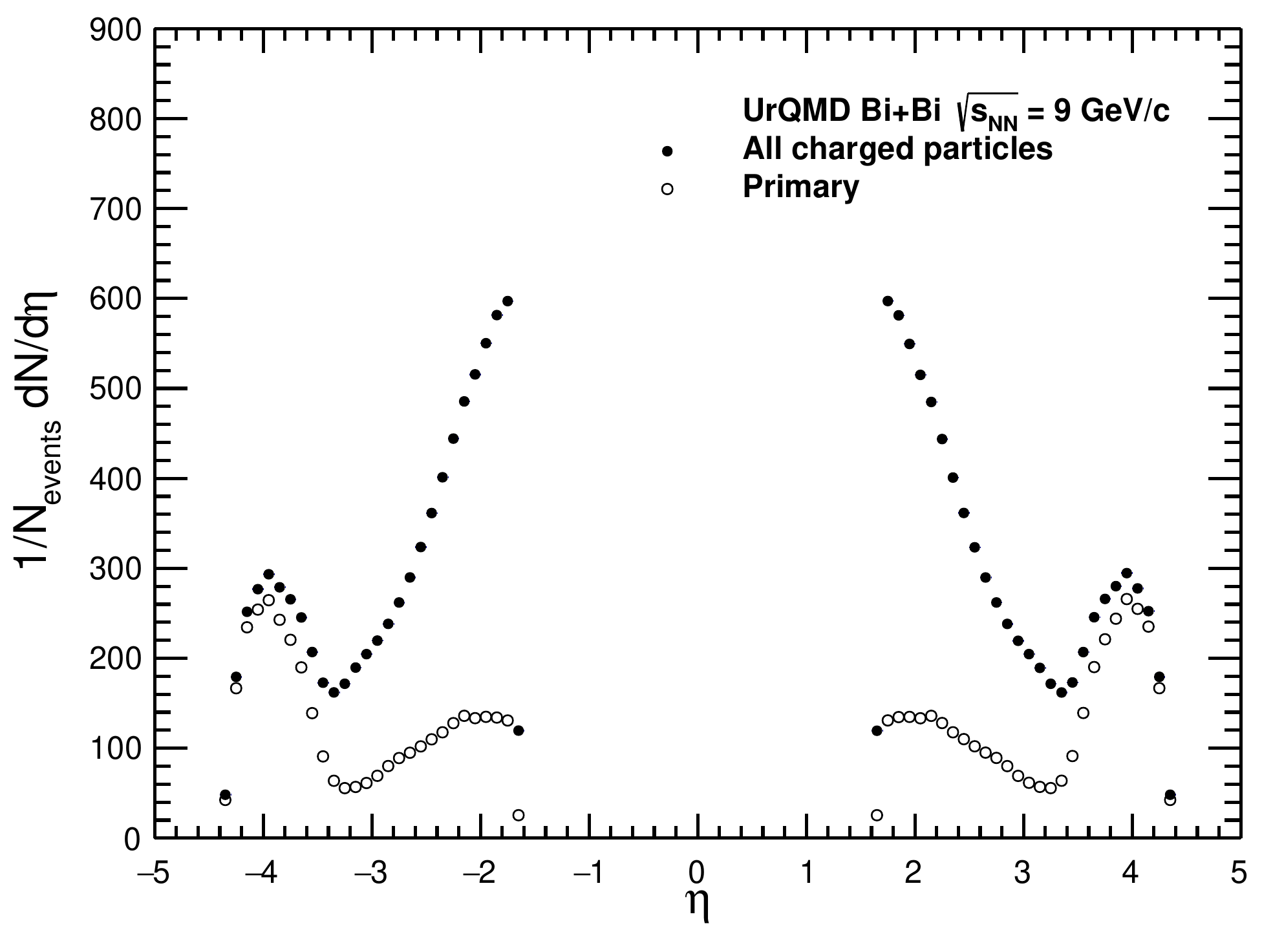}
\caption{Pseudorapidity coverages of BeBe detector, for all charged particles and primary particles.}
\label{fig:BBeta}
\end{figure}

The BeBe detector is planned to be made of two arrays of BC-404 plastic scintillator counters located 2 meters away from the MPD-NICA interaction point (IP), at opposite sides, see Fig.~\ref{fig:BBGeometryIP}. Each of the arrays will consist of 80 individual 1 cm thick cells wrapped with layers of mylar and Tyvek. BeBe will cover a pseudorapidity range of $1.68 < |\eta|< 4.36$, see Table~\ref{Tab:eta} and Fig.~\ref{fig:BBeta}.  The considered geometry for BeBe is similar to the one used for the VZERO-ALICE ~\cite{Abbas:2013taa} detector during Run 1 and 2 of the LHC. 

It is expected that the light produced in the sensitive material will be collected by Silicon PhotoMultipliers (SiPMs) coupled directly to each individual plastic scintillator cell. As it was reported in~\cite{Alvarado:2018gbb}, the SiPMs photosensor may provide an excellent intrinsic time resolution for the detector of the order of tens of picoseconds. Nevertheless, other methods to extract the light from the plastic scintillator are under investigation, such as optical fibers coupled to SiPMs. 


\begin{table}[hbt!]
\centering
\begin{tabular}{|c|c|c|c|}
\hline
Ring & \multicolumn{1}{c|}{$\eta$} & \multicolumn{1}{c|}{R$_{min}$} & \multicolumn{1}{c|}{R$_{max}$}\\ \hline
1         & 3.87 - 4.36    & 5.1    & 8.3            \\ \hline
2         & 3.31 - 3.87    & 8.5     & 14.5            \\ \hline
3         & 2.84 - 3.31    & 14.7     & 23.4          \\ \hline
4         & 2.26 - 2.84    & 23.6    & 42           \\ \hline
5         & 1.68 - 2.26    & 42.2    & 76.63           \\ \hline
\end{tabular}
\caption{Pseudorapidity and BeBe rings dimensions (R$_{min}$ is the minimum radius of the ring and R$_{max}$ is the maximum radius shown in cm).}
\label{Tab:eta}
\end{table}

\clearpage
\section{Simulation of BeBe cell prototype with Geant-4}
\label{sec:BeBe_corners}

To estimate the dispersion in the optical photon arrival time to the effective photon detection area~(EA), simulations with GEANT-4 v.10.06 toolkit software~\cite{AGOSTINELLI2003250} were performed. The simulation considered cells of 1.5~cm thickness taking the BC404 plastic scintillator as sensitive material ~\cite{bc404data,bc404thesis}. This time dispersion is considered as the intrinsic time resolution~(ITR) of the simulated BeBe cell. Three different cell sizes were simulated accordingly with the dimensions of BeBe rings 1, 3, and 5, see Table~\ref{Tab:eta}. The simulated EA had an area of $6 \times 6~mm^2$ and was coupled directly to the BeBe cell for two configurations: at the center and at the superior left corner of the cell. One thousand Monte Carlo events were generated with 1 ~GeV muons striking the BeBe cell in the center, in the corner, and randomly distributed over the entire frontal area of the cell. In total, six configurations per cell were simulated. For this study, two boundaries surface were considered:
\begin{itemize}
\item Scintillator-environment surface: It was simulated 95\% reflective. A polished plastic was considered.
\item Scintillator-EA surface: It was simulated with 100\% absorption, in order to avoid double counting of optical photons arriving at the EA. 

\end{itemize}

\begin{figure}[!hbt]
\centering\includegraphics[width=0.6\linewidth]{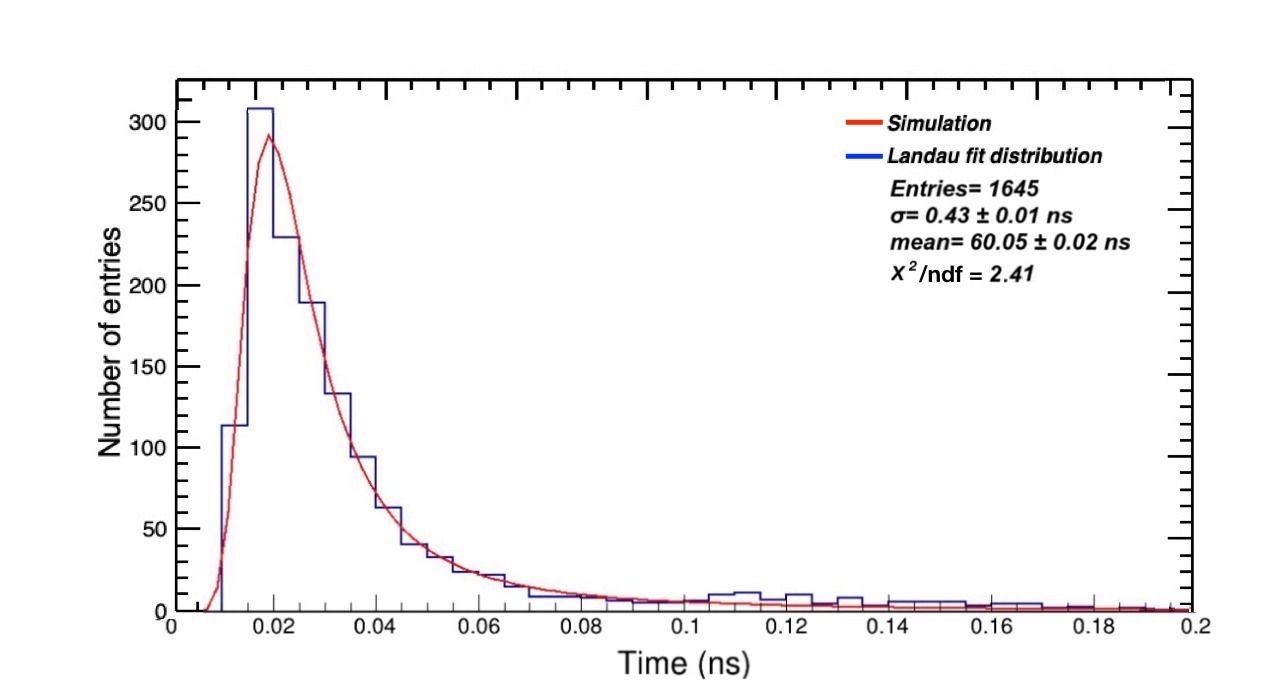}
\caption{Arrival time optical photon distribution to the EA for one event.}
\label{fig:landaudist}
\end{figure}

\begin{figure}[!hbt]
\centering\includegraphics[width=1\linewidth]{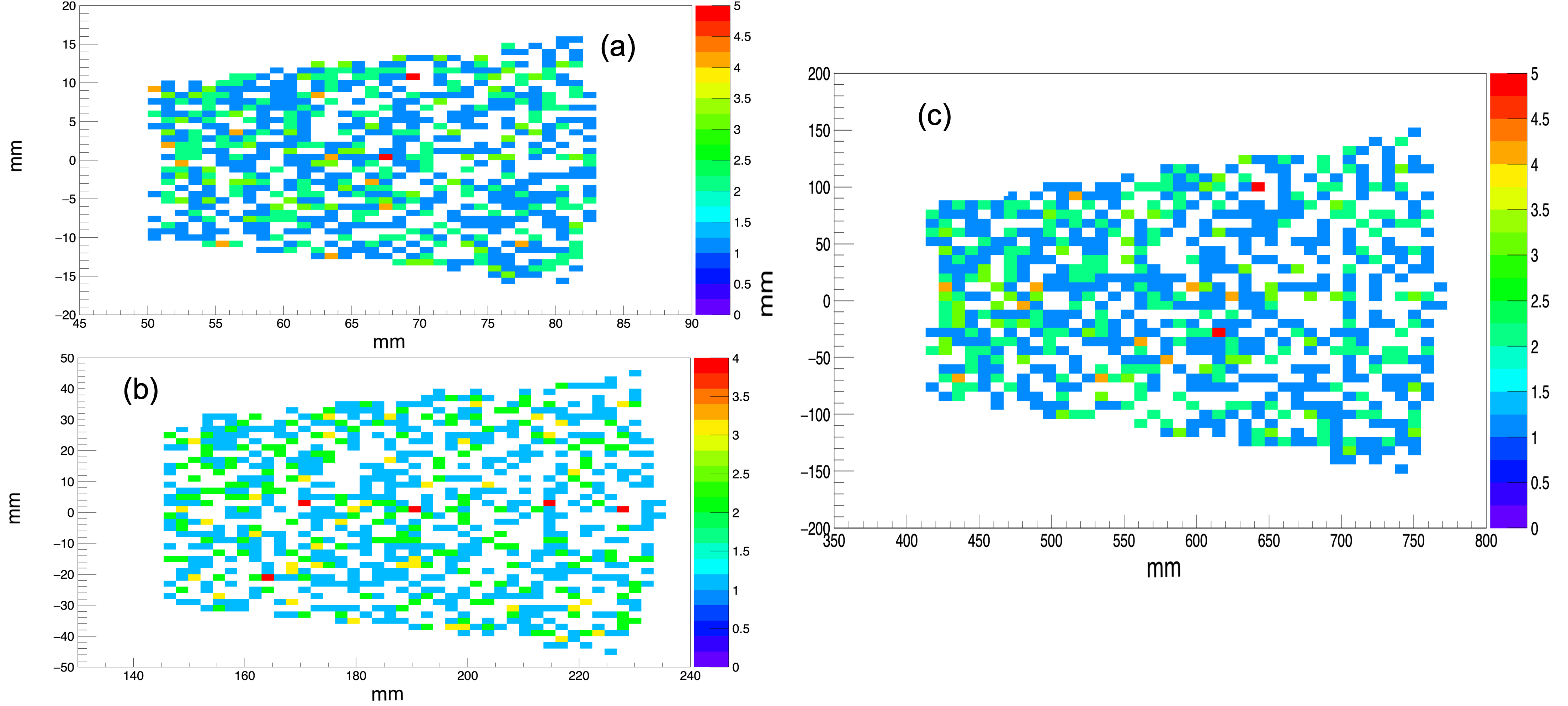}
\caption{Hits distribution of the simulated particles in the BeBe cells of ring 1 (a), ring 3 (b), and ring 5 (c).}
\label{fig:BeBeHits}
\end{figure}

Event by event we plotted the optical photon arrival time (OPAT) to the EA. In Fig. ~\ref{fig:landaudist} an example of an OPAT distribution for one single event is shown. Fitting the OPAT distributions with a Landau function, we estimated numerically the mean value, using the ROOT tools. Also, from the ROOT tools, we estimated the most probable value (MPV) of the optical photon arrival time per event. To perform this analysis, we selected only those events where the fit to the OPAT distribution gives a $chi^2/ndf < 2$ and the difference between the mean value from the fit and the mean value from the OPAT distribution is less than 20\%. Fitting the distribution constructed with all the MPV extracted from the Landau fits to the OPAT distributions, we estimated the intrinsic time resolution (ITR). Three cases were studied: a) the generated particle hits the center of the simulated cell, b) the generated particle hits the corner of the simulated cell and c) the generated particle hits randomly the surface of the simulated BeBe cell, see Fig. ~\ref{fig:BeBeHits} 


\begin{figure}[!hbt]
\centering\includegraphics[width=1\linewidth]{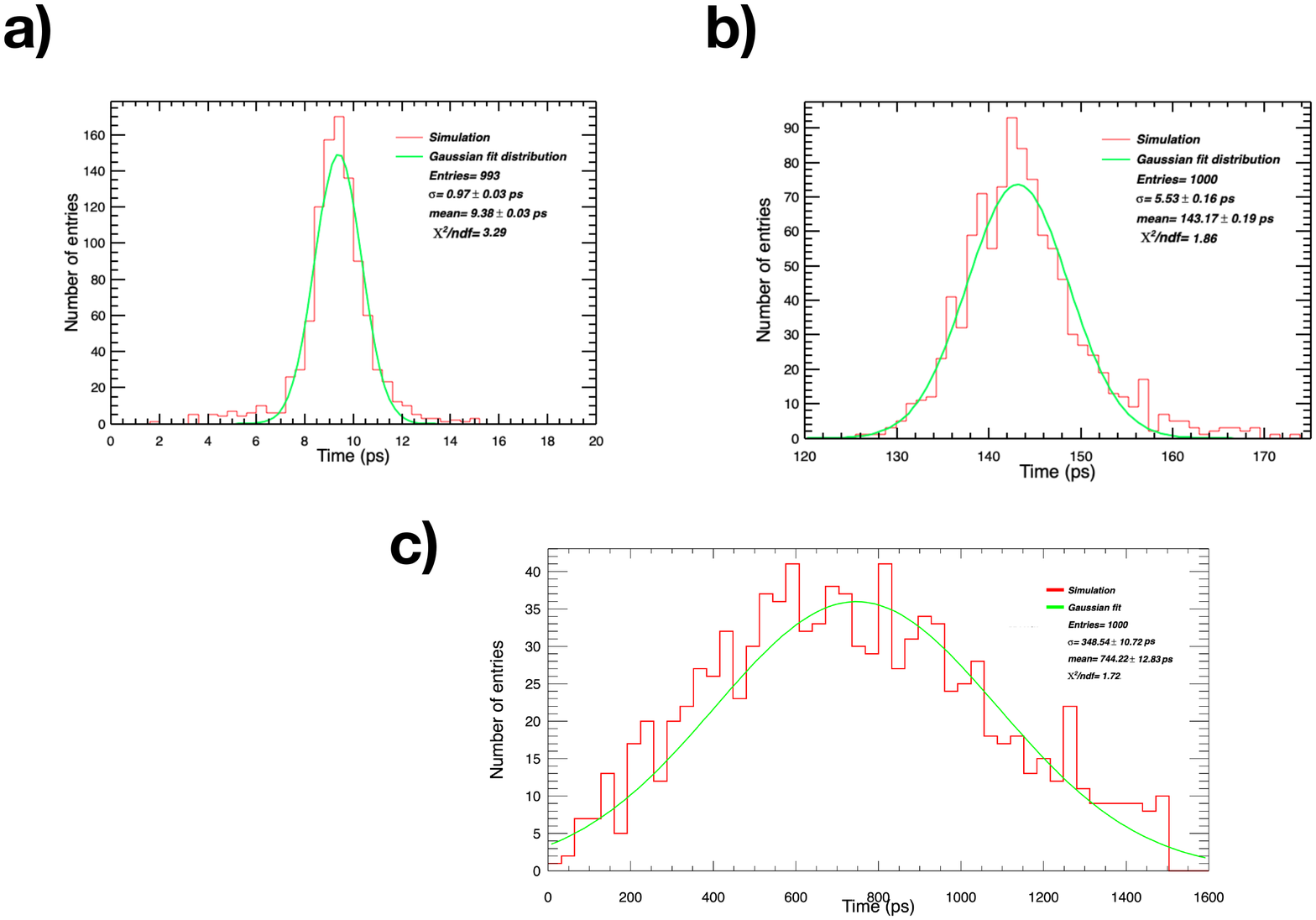}
\caption{Mean arrival optical photon distribution for one EA at the center and muons interaction at a) center, b) corner, and c) random.}
\label{fig:gaussiandist}
\end{figure}

For a) and b), the generated particles hit the simulated cell in the same position and the MPV plot behaves as Gaussian, see Fig. ~\ref{fig:gaussiandist}-a,b. For c), 
the distribution of hits shown in Fig. ~\ref{fig:BeBeHits} is almost uniform with some blank regions and the MPV plot behaves as a wide Gaussian distribution, see Fig. 5-c.


\begin{figure}[!hbt]
\centering\includegraphics[width=0.65\linewidth]{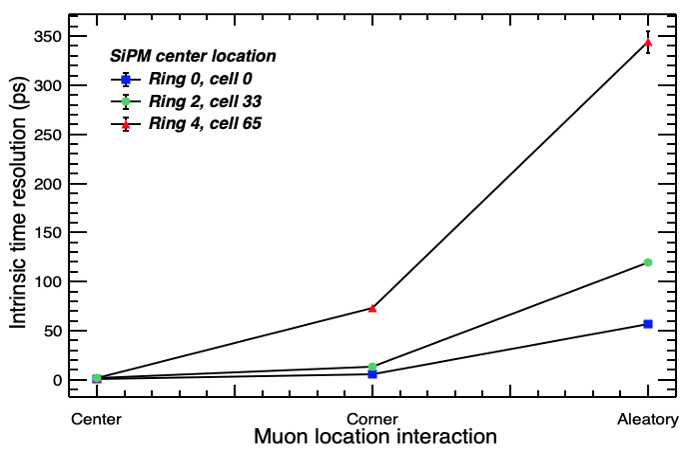}
\centering\includegraphics[width=0.65\linewidth]{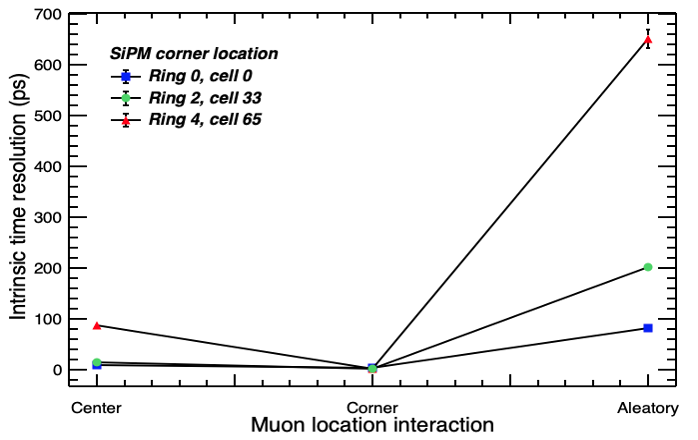}
\caption{Intrinsic time resolution values for the three muon hit locations and both EA positions.}
\label{fig:geantresults}
\end{figure}

The ITR is not constant and it depends on the hit location of the generated particle into the BeBe cell and the location of the EA (SiPM). The ITR for arbitrary cells of the rings 1, 3, and 5 is of the order of 6, 15, and 82~ps for one EA located in the center of the cell and 18, 20, 60~ps for one EA located in the corner of the cell respectively. 
The ITR presented here is independent of the electronics and data acquisition system and it only depends on the geometry of the scintillator, number of photosensors, specie and energy of incident particles.

\clearpage
\section{Laboratory measurements of time resolution}
\label{sec:Measurements}

\subsection{Experimental setup}
To determine the time resolution of the BeBe cells, two prototypes of the configuration array of the system were studied using secondary particles of cosmic rays as a radiation source. Each prototype consists of an array of photodetectors coupled to ultra-fast plastic scintillator BC-404 from Saint-Gobain Crystals wrapped with one layer of Tyvek and two layers of Mylar. The plastic scintillator sizes are shown in Table~\ref{Tab:Cells}, they are 2~cm thick and we labeled the smallest as P1 and the largest as P2. We selected these two scintillator prototypes because their dimensions are closed to the dimensions of the rings 4 and 5 of the BeBe used in the simulation sections~\ref{sec:BeBe_corners} and \ref{sec:Simulations}, (see table~\ref{Tab:eta}).

\begin{figure}[hbt!]
\centering
\includegraphics[scale=0.35]{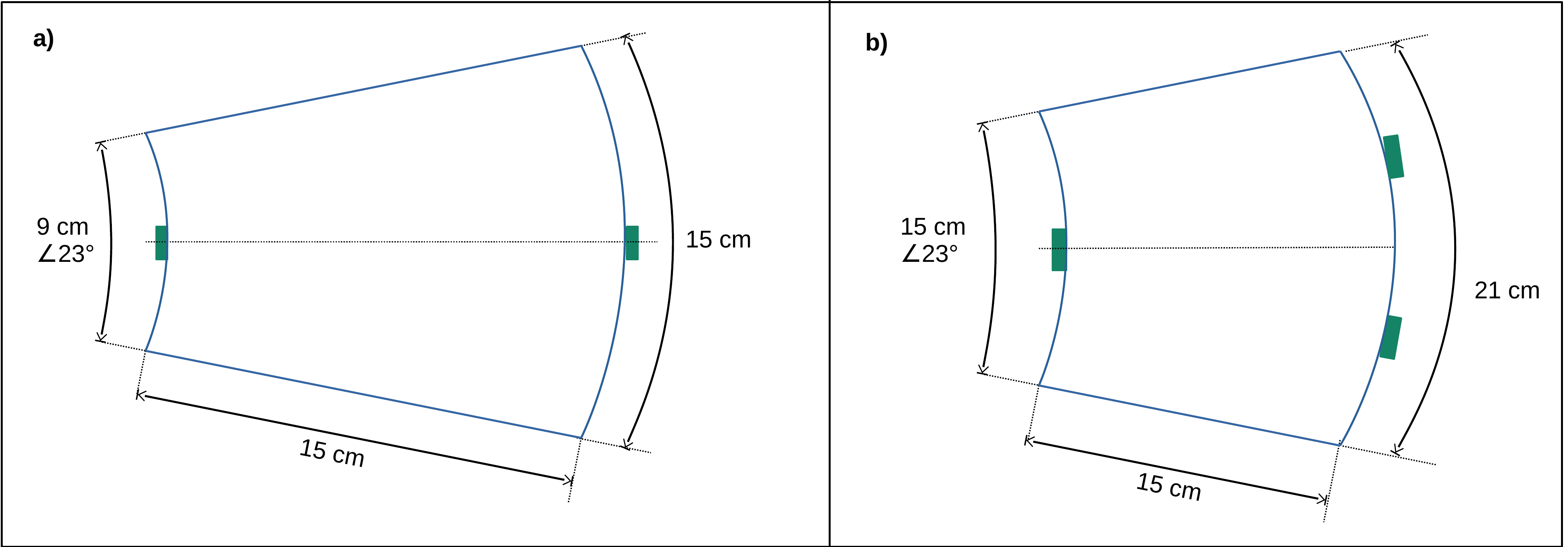} 
\caption{Diagram of the prototypes for the two configurations used in the measurements, the position of counters is marked in green. a)Two photosensors attached to the smallest prototype~(P1). The configuration with one inner/outer sensor consists of one sensor in the center of the smallest/largest face. b)Three sensors in the largest cell~(P2).}
\label{fig:Prototypes}
\end{figure}

The photosensors were coupled to the cells in 4 different ways: i. one sensor in the center of the inner~(or shortest) lateral face ii. one sensor in the center of the outer~(or largest) lateral face. iii. two sensors in the center of each lateral face~(the combination of the first two)  iv. three sensors with one in the center of the inner lateral face and two equally distributed in the outer lateral face. The third and fourth configurations are shown in Fig.~\ref{fig:Prototypes}.-a) and -b), respectively.

Different photodetectors were used to collect the scintillation light from the disk cells to compare the effect of their different effective area: 3 Silicon photomultipliers~(SiPMs) from Hamamatsu Photonics~(HPK) S13360-3050CS, 2 SiPMs SensL MICROFC-60035-SMT-TR1, and 2 photomultiplier tubes~(PMTs) Hamamatsu H5783~(identified as PMT-HPK). The SiPM characteristics and operation conditions are summarized in Table~\ref{Tab:SiPM} and its polarization circuit is described in section~\ref{sec:Electronics}. The PMTs-HPK input voltage was set to 14.0~V and control voltage at 0.75~V. The characteristic values from each photodetector signal are summarized in Table~\ref{Tab:Signals}.

The SiPM-HPK/-SensL/ PMT-HPK has a $0.3\times0.3$/ $0.6\times0.6$/ $1\times1$~$cm^{2}$ sensitive area, while the inner lateral face of the P1 is $9\times 2 ~cm^{2}$, each of the sensors cannot collect more than 0.5\%/ 2.0\%/ 5.5\% of incoming light. The collection of incoming light decreases by 40\% for the outer lateral face of P1~($15\times 2 ~cm^{2}$, equal dimensions of the inner lateral face of P2) and by 67\% for the outer lateral face of P2~($21 \times 2~cm^{2}$).

\begin{figure}[hbt!]
\centering
\includegraphics[scale=0.4]{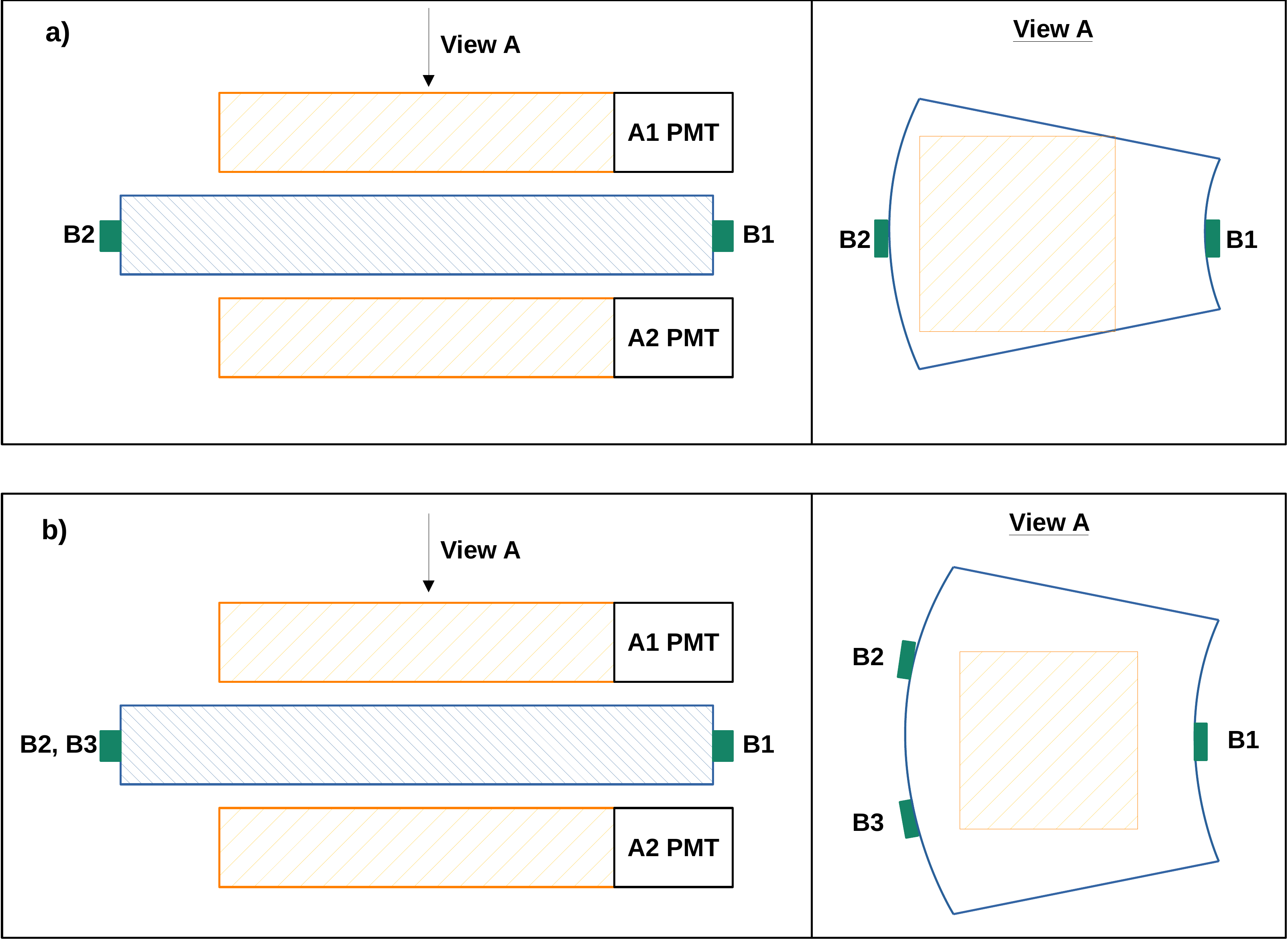} 
\caption{Schematic layout of the displacement of the trigger counters~(in yellow), the prototypes~(in blue), and the position of the photosensors~(in green), with two views, horizontal and top. a) Two sensors attached to the smallest cell~(P1) and b) Three photodetectors attached to the largest prototype~(P2).}
\label{fig:Experimental_Setup}
\end{figure}

The experimental setup is shown in Fig.~\ref{fig:Experimental_Setup}, the prototypes~(in blue) were placed between the trigger counters~(in yellow), which provided the start signal for data readout. The trigger counters were made of two BC404 scintillator plates of $10\times10\times2~cm^{3}$ and two PMTs Hamamatsu H5783. Each trigger counter consists of a PMT coupled to the plate in the center of one of the lateral faces. Light from the scintillators was detected by the PMTs at one end of the bar, and the coincidence of the signals from the two PMTs crossing the threshold level of 10~mV was the external trigger for
the CAEN digitizer DT5720B. 

\begin{figure}[hbt!]
\centering
\includegraphics[scale=0.6]{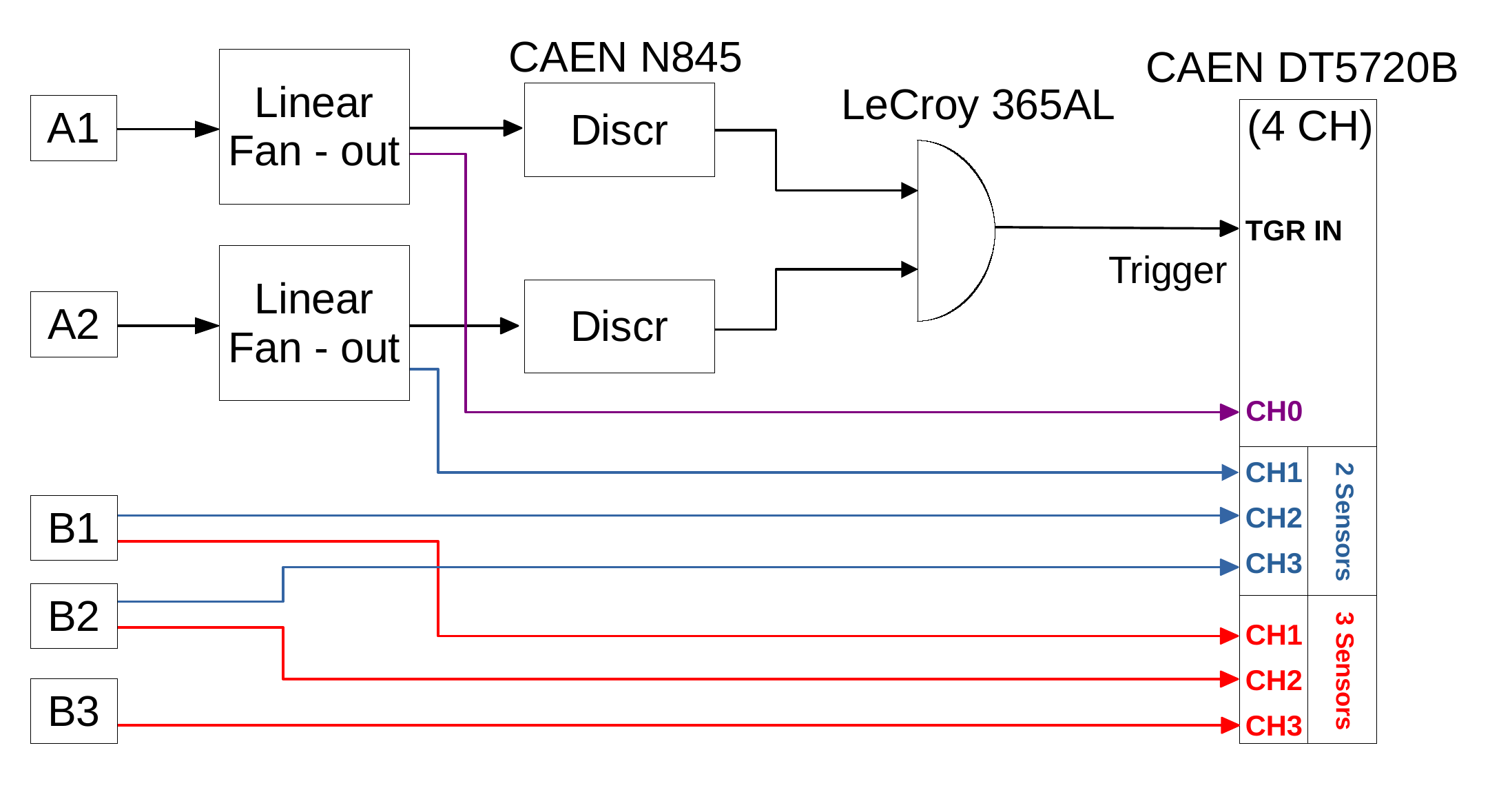}
\caption{Readout electronics scheme used to digitize pulses. The black lines correspond to the PMTs’ signals (A's) from the trigger counters, their coincidence generates the trigger of the system. The A1 signal (purple line) was used as a reference for all configurations. The configuration for 2 and 3 photodetectors (B’s signals) attached to the prototype is shown in blue and red lines, respectively. The configuration for one inner/outer sensor is the same as the configuration for two sensors considering only one signal from B1/B2.}
\label{fig:DAQ}
\end{figure}

Because the CAEN digitizer DT5720B has 4 acquisition channels, it was necessary to change the configuration in the readout when 2 or 3 photomultipliers were attached to the prototypes. The readout electronics is schematically shown in Fig.~\ref{fig:DAQ}, where the configuration with two/three photosensors attached is shown in blue/red. The sampling frequency of the digitizer was set to the highest level (250~MHz, 1024-bin-long waveform, 4~ns time bin). The recorded data sets were analyzed offline on an event-by-event basis.

\begin{table}[hbt!]
\centering
\begin{tabular}{|c|c|c|c|}
\hline
Cell & \multicolumn{1}{c|}{$\eta$} & \multicolumn{1}{c|}{R$_{min}$} & \multicolumn{1}{c|}{R$_{max}$}\\ \hline \hline
P1         & 2.39 - 2.90    & 22    & 37            \\ \hline
P2         & 2.06 - 2.39    & 37     & 52            \\ \hline
\end{tabular}
\caption
{Dimensions of the scintillators used for the laboratory measurements.
The cells correspond to rings with minimum~(R$_{min}$) and maximum~(R$_{max}$) radius given in cm. Each cell with a width of 2~cm and an angular coverage of $23^{o}$. The rings cover the pseudorapidity shown with the center of the rings at 200 cm from the I.P.}
\label{Tab:Cells}
\end{table}

\begin{table}[tb!]
\centering
\begin{tabular}{|c|c|c|c|c|}
\hline
SiPM & $V_{Br}$~(V) & $V_{Bias}$~(V) & Microcell size & Microcells   \\ \hline \hline
SensL         & $25.7\pm0.5$    & 28.0 \& 30.5     &  35~$\mu m$ &18980 \\ \hline
HPK         & $52.7\pm0.5$   & 54.1 \& 55.1    &  50~$\mu m$ & 3600          \\ \hline
\end{tabular}
\caption
{Key SiPM characteristics and operating conditions. We considered two different $V_{bias}$.}
\label{Tab:SiPM}
\end{table}

\begin{table}[tb!]
\centering
\begin{tabular}{|c|c|c|c|c|}
\hline
Prototype & Photodetector & Rise Time~(ns) & Amplitude~(mV) & Signal width~(ns) \\ \hline \hline
\multirow{5}{*}{P1}  
& PMT-HPK                     & $8.7\pm2$   &    $122.9\pm21.8$ & $17.8\pm2$   \\ \cline{2-5}
 & SensL ($V_{Bias}=28.0~V$)  & $20\pm2$   &    $33.6\pm5.1$   & $170\pm2$   \\ \cline{2-5}
 & SensL  ($V_{Bias}=30.5~V$) & $21\pm2$   &    $96.7\pm16.7$   & $172\pm2$   \\ \cline{2-5}
 & HPK   ($V_{Bias}=54.1~V$)  & $17\pm2$   &    $22.9\pm3.5$   & $51\pm2$   \\ \cline{2-5}
 & HPK  ($V_{Bias}=55.1~V$)   & $17\pm2$   &    $37.0\pm5.6$   & $17\pm2$   \\ \hline \hline
 
\multirow{5}{*}{P2} 
& PMT-HPK                       & $10\pm2$   &    $118.0\pm25.6$   & $18\pm2$   \\ \cline{2-5}
 & SensL    ($V_{Bias}=28.0~V$) & $21\pm2$   &    $32.6\pm5.2$   & $174\pm2$   \\ \cline{2-5}
  & SensL   ($V_{Bias}=30.5~V$) & $22\pm2$   &    $80.7\pm13.6$   & $173\pm2$   \\ \cline{2-5}
 & HPK  ($V_{Bias}=54.1~V$)     & $17\pm2$   &    $19.9\pm3.1$   & $51\pm2$   \\ \cline{2-5}
 & HPK  ($V_{Bias}=55.1~V$)     & $16\pm2$   &    $31.1\pm5.0$   & $52\pm2$   \\ 
 \hline

\end{tabular}
\caption
{Photodetector signal properties. The rise time was calculated with the difference between the times when the signal front crosses the 10\% and 90\% of the signal height. The signal width corresponds to the FWHM.}
\label{Tab:Signals}
\end{table}

\clearpage
\subsection{Data analysis}
\label{sec:DataAnalysis}
The acquired waveform signals provided by the CAEN digitizer were analyzed with the method discussed in detail in~\cite{Torres:2019ntb}. The general aspects of this method are described in the next paragraphs. 
We expected that the time resolution of each BeBe prototype depends on the volume (P2 is bigger than P1) due to internal light losses due to multiple reflections and the reduced amount of light arriving in every photodetector. 

 The selection criteria for each event consist of a coincidence between the trigger counters and the signal in all the photodetectors placed in the prototype crossing the threshold level at 10~mV/ 14~mV/ 100~mV for the SiPM-HPK/ SensL/ PMT-HPK. 

 As reported in~\cite{VINKE2009188, Torres:2019ntb}, the optimal time resolution for SiPMs has been obtained with a digital Constant Fraction Discriminator~(dCFD) to measure the arrival time~(t0) of the signals~\cite{BARDELLI2004480} using a  30\% threshold relative to the maximum value of the pulse. Following a similar procedure as the one described in~\cite{Torres:2019ntb}, we found that a photomultiplier tube reaches its optimal time resolution considering the dCFD method with a threshold level equal to the 50\% fraction of the pulse maximum.


We performed two kinds of time resolution analyses, one of them was to select events from the fastest signal from the two or three photosensors attached. The second approach used the average arrival time of the signals. Each analysis showed similar results and we present the results for the fastest arrival time of the signal.

\begin{figure}[hbt!]
\centering
\includegraphics[scale=0.32]{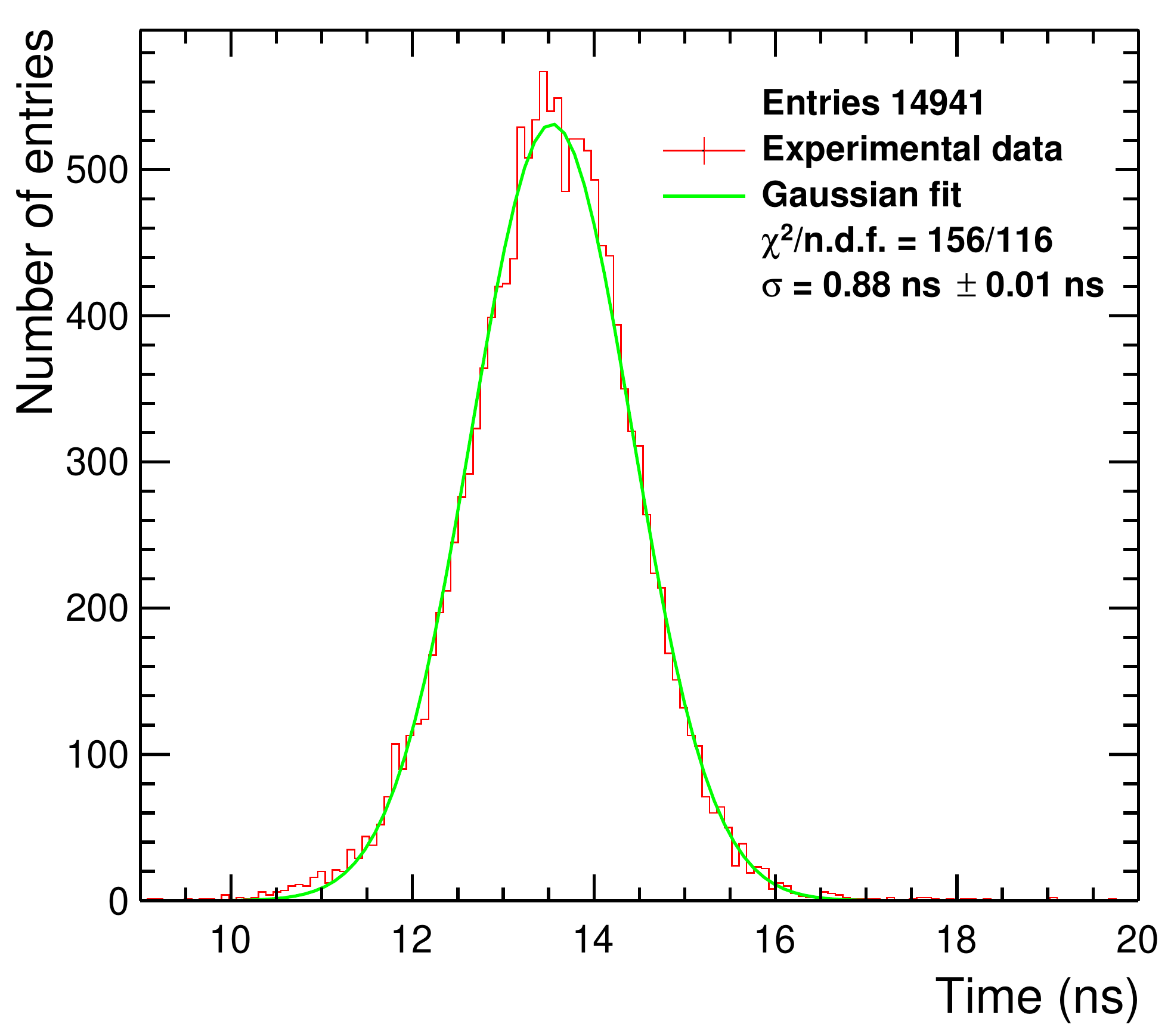} 
\caption{Time difference between the arrival time of the signal in the trigger counter and the fast arrival time of the signal for the smaller prototype cell~(P1) with 3 SiPMs-HPK attached ($V_{bias}$ = 54.1~V).}
\label{fig:sigma_AB}
\end{figure}

The cosmic rays' impact time in the prototype~($t_{B}$) is the fastest arrival time considering all the incoming signals from the photodetectors attached.
For the trigger counters A1 and A2, the arrival time is $t0_{A1}$ and $t0_{A2}$, respectively. Then, the difference between the times $(t0_{A1} - t_{B})$ was obtained, an example of the difference is shown in Fig.~\ref{fig:sigma_AB} for 3 SiPMs-HPK attached to P1. We performed a Gaussian fit over all the range of the experimental data, where the standard deviation corresponds to $\sigma_{A1,B}$. Similar procedures were followed with the distributions obtained for the other configurations. 

The time resolution of the trigger counters was obtained considering the difference between the times $(t0_{A1} - t0_{A2})$ and considering that the standard deviation corresponds to $\sigma_{A1,A2}$, in addition, it is taken into account that both counters A1 and A2 have the same time resolution since both are identical and work under the same conditions, then, $\sigma_{A1}=\sigma_{A2}= \frac{\sigma_{A1,A2}}{\sqrt{2}}$ and equal to $0.58\pm0.02~ns$. 
As a result, the time resolution of the prototypes $\sigma_{B}$ was obtained considering that $\sigma_{A1,B}^2 = \sigma_{A1}^2 + \sigma_{B}^2$, then $\sigma_{B}=(\sigma_{A1,B}^2 - \sigma_{A1}^2)^{1/2}$.

\begin{figure}[hbt!]
\centering
\includegraphics[width=1.\textwidth]{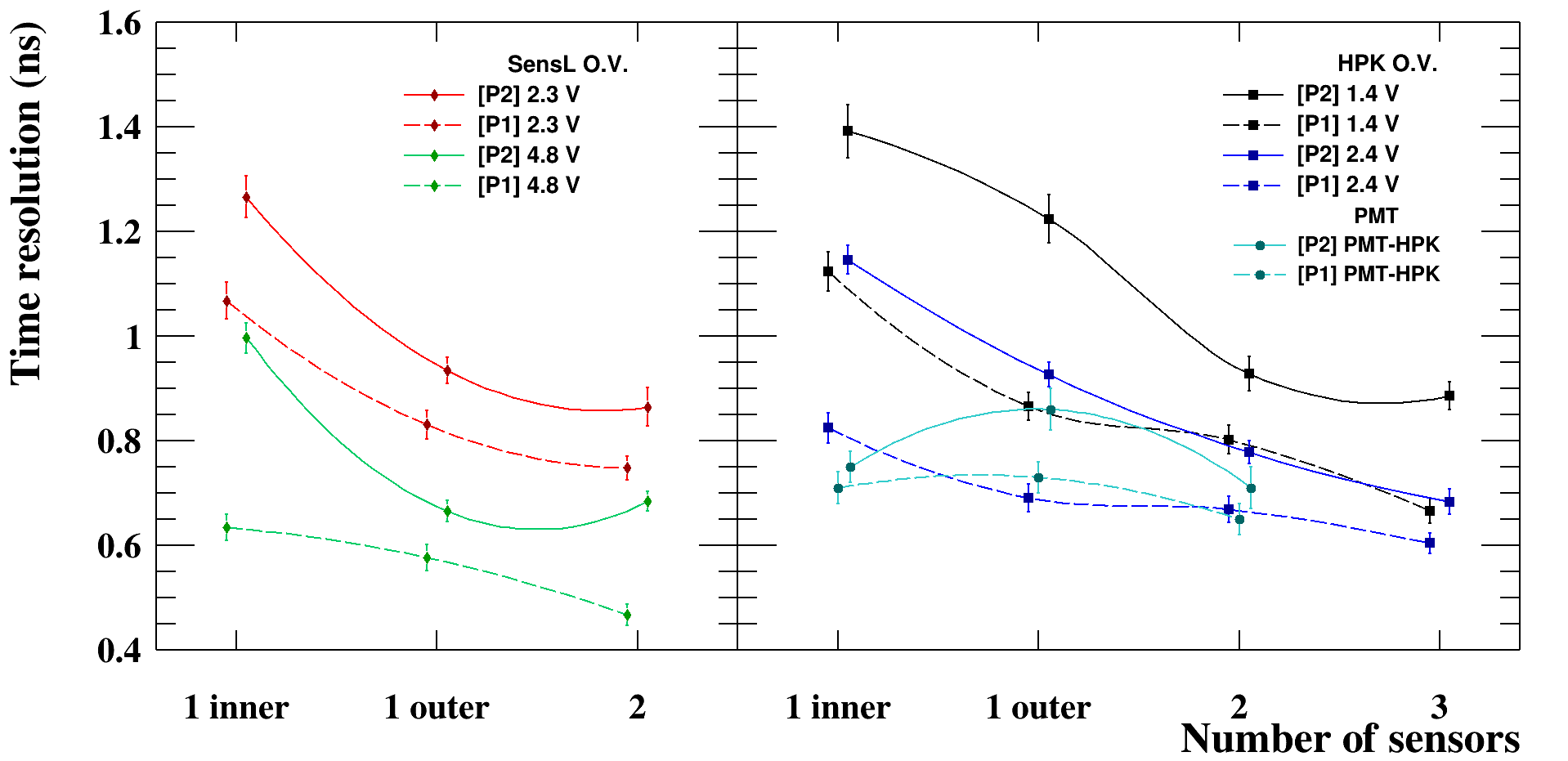}
\caption{Time resolution of the counters obtained by each configuration. Left) Time resolution for the SensL considering operating voltages of 28 and 30.5 V, which corresponds to the O.V. of 2.3 $\pm$ 0.5 V and 4.8 $\pm$ 0.5 V, respectively. Right) The time resolution for the PMT and SiPM HPK. The different results for the SiPM-HPK corresponds to two operating voltages of 54.1 and 55.1~V, corresponding to the O.V. of 1.4 $\pm$ 0.5 V and 2.4 $\pm$ 0.5 V, respectively.}
\label{fig:Plot_TR}
\end{figure}

Figure~\ref{fig:Plot_TR} shows the time resolution of the counters obtained by each configuration. For one counter, we determined the time resolution of one sensor in the inner lateral face~(1~inner) and one counter in the outer lateral face~(1~outer). For the PMTs-HPK attached to P1 and P2, the time resolution reached by the 3 different configurations is similar, in the case of P1 it ranges from 0.65$\pm$0.03~ns to 0.71$\pm$0.03~ns and for P2 it ranges from 0.71$\pm$0.04~ns to 0.75$\pm$0.03~ns.

The photodetection efficiency~(PDE) of the SiPMs HPK\footnote{HPK datasheet:\url{https://www.hamamatsu.com/content/dam/hamamatsu-photonics/sites/documents/99_SALES_LIBRARY/ssd/s13360_series_kapd1052e.pdf}} and SensL\footnote{SensL datasheet:\url{ https://www.onsemi.com/pdf/datasheet/microc-series-d.pdf}} increases as a function of the overvoltage. We considered two O.V. values for each SiPM and we observed that by increasing the overvoltage the time resolution reaches lower values. For one SiPM, the outer counter reaches the best time resolution, this was expected since the trigger counter was placed closer to the outer face of the prototype, as in this position the counter area is within the prototype dimensions. 

The time resolution for the SiPM-SensL attached to P1 goes from
0.75$\pm$0.02 to 1.07$\pm$0.04~ns and from 0.47$\pm$0.02 to 0.63$\pm$0.03~ns for the $V_{bias}=28.0~V$ and $30.5~V$, respectively. For P2, from 
0.86$\pm$0.4 to 1.27$\pm$0.04~ns and from 0.67$\pm$0.02 to 1.00$\pm$0.03~ns, for the $V_{bias}=28.0~V$ and $30.5~V$, respectively.

For the SiPM-HPK the values of the time resolution for P1 go from
0.67$\pm$0.03 to 1.12$\pm$0.04~ns and 0.60$\pm$0.03 to 0.82$\pm$0.02~ns, for $V_{bias}$ of 54.1~V and 55.1~V, respectively. 
For P2, from 0.89$\pm$0.03 to 1.39$\pm$0.05~ns and from 0.68$\pm$0.3 to 1.14$\pm$0.3~ns, for $V_{bias}$ of 54.1~V and 55.1~V, respectively.

\begin{figure}[hbt!]
\centering
\includegraphics[width=0.85\textwidth]{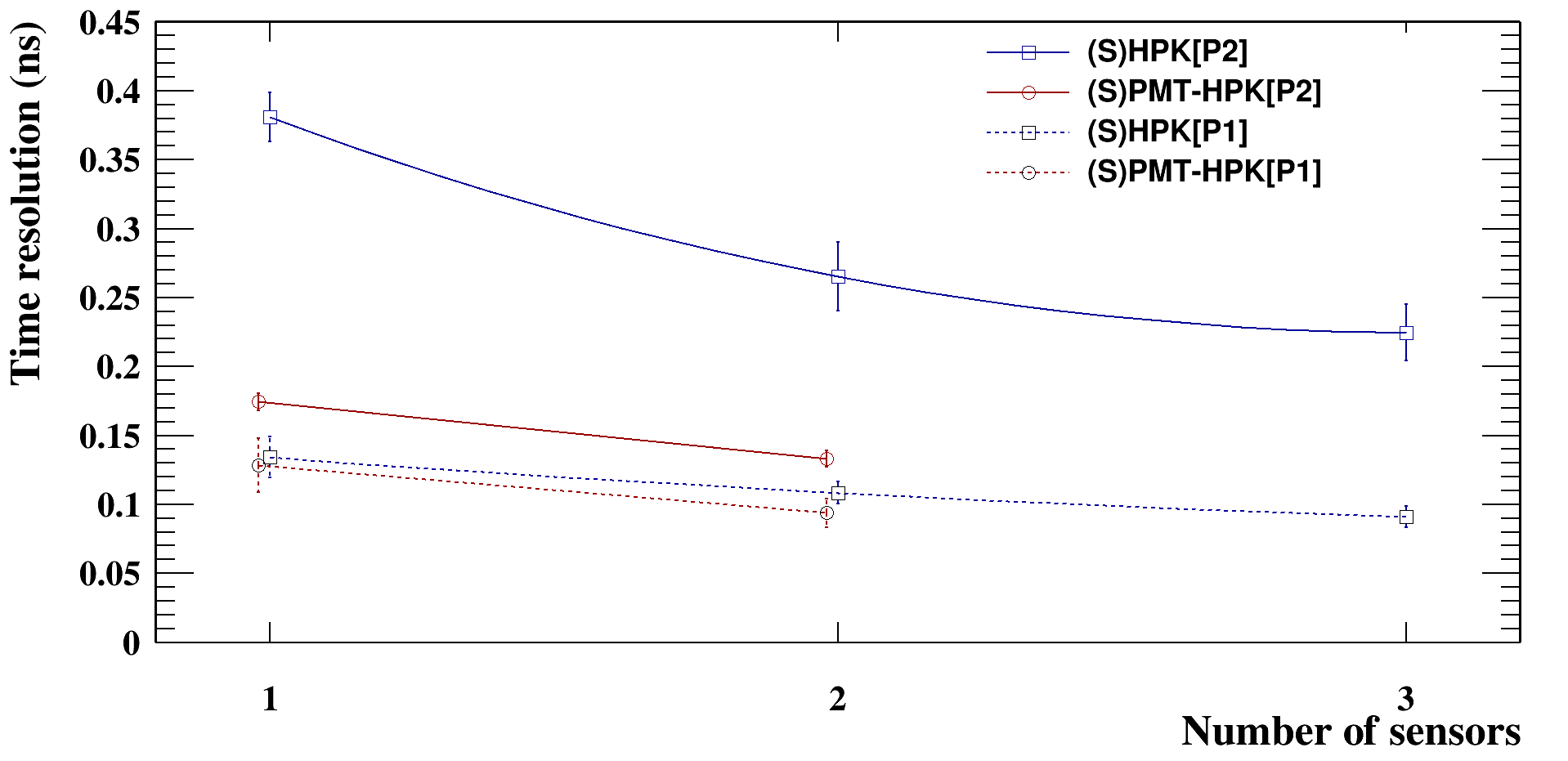}
\caption{Intrinsic time resolution of the prototypes P1 and P2 obtained with a Geant-4 simulation(S). The dimension of the EA was simulated considering the effective area of the photosensors. The score number 1 corresponds to one EA in the inner lateral face.}
\label{fig:Fast_TR}
\end{figure}

The ITR of the prototypes of the different configurations that involve PMTs and SiPMs Hamamatsu was estimated with a Geant-4 v10.06 simulation. For this analysis, the geometry scintillator was simulated as it is shown in Fig.~\ref{fig:Prototypes}. The dimension of the EA was simulated considering the effective area of the PMTs and the SiPMs Hamamatsu and they were coupled as it is shown in Fig.~\ref{fig:Experimental_Setup}. The surrounding environment, the optical boundaries between scintillator-environment, and scintillator-EA surfaces of the simulated prototype were considered the same as mentioned in section 3. Finally, muons were considered as incident particles to the scintillators' frontal surface area and following the energy spectrum distribution of the cosmic rays on Earth's surface~\cite{muons}. The ITR values are shown in Fig.~\ref{fig:Fast_TR}.
Comparing the laboratory results against the simulation we can observe an evident difference due to particular features not taken into account in the simulation, for instance, the PDE of the SiPMs, light losses, and electronic noise in the laboratory setup. 
It is important to highlight that the simulation results in Geant4, are of great importance to identify the expected number of photons to collect in the detection areas. In this way, the expected deposited charge in the SiPM can be estimated.

As we expected, the three sensor configuration gives us the best time resolution result, for both prototypes.  
Also, considering the dimensions of the prototypes, we estimated a time resolution of around 1~ns is, for this first study, a good initial one. This study is relevant to figuring out including more sensors to obtain better time resolution and by decreasing the width of the cells. 

\clearpage
\subsection{Readout electronics}
\label{sec:Electronics}

\begin{figure}[hb!]
	\centering
	\includegraphics[width=0.9\linewidth]{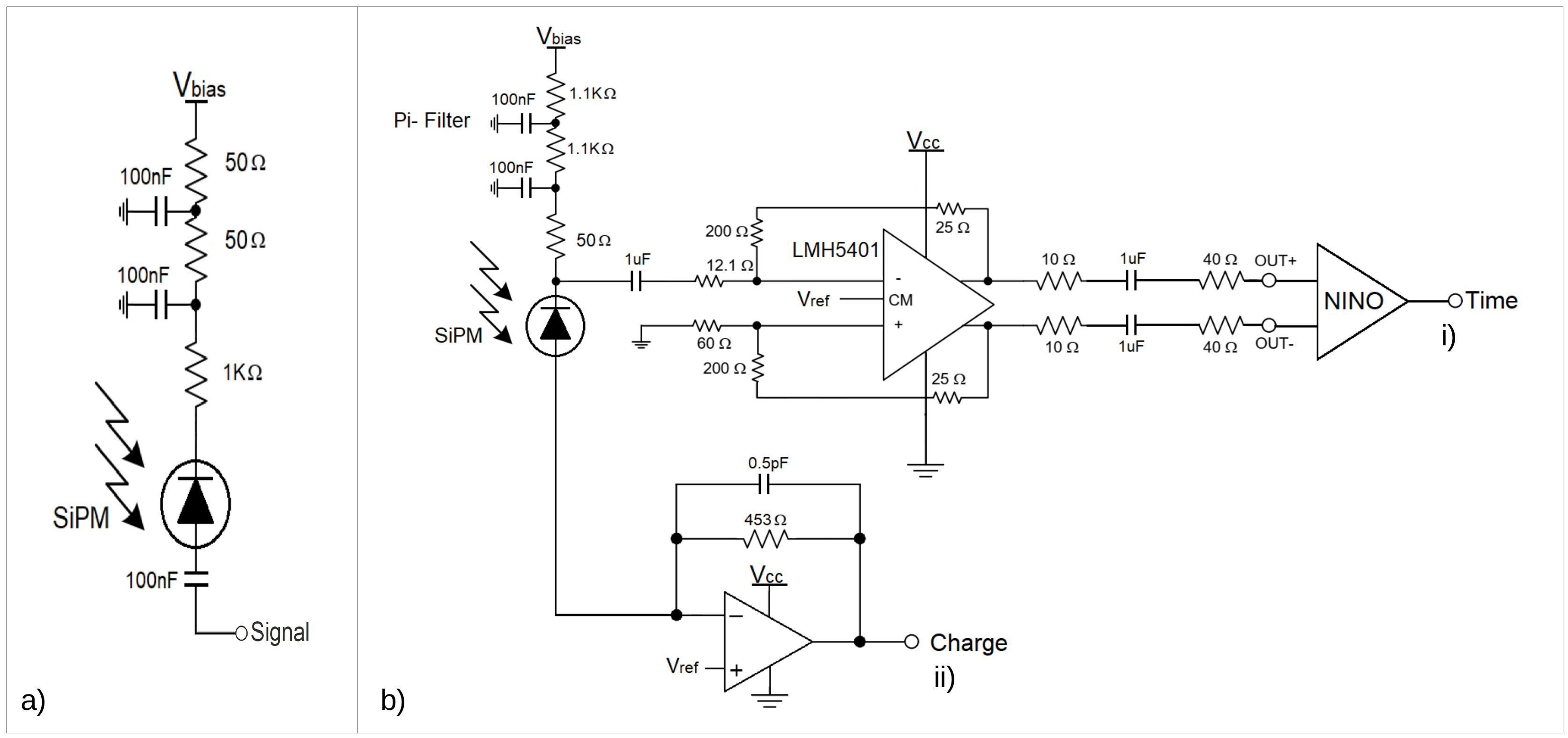}
	\caption[]{BeBe Front-end schematics. b.i) Readout using single-ended to differential amplifier LMH5401 with NINO chip, for timing information and triggering, b.ii) Charge integration with trans-impedance amplifier OPA858 for energy information.}
	\label{fig:feschematics}
\end{figure}

The time resolution analysis considering different photosensors showed a better time resolution for the PMT-HPK than the one considering SiPMs. The PDE and gain of the PMT are higher than the SiPM.  The polarization circuit of the SiPM as is shown in Fig~.\ref{fig:feschematics}.a) does not include an amplification stage. For that reason, as is shown in Fig.~\ref{fig:feschematics}.b) we want to introduce an amplification stage to achieve a better time resolution. 

As BeBe is planned as a forward detector in MPD, information about deposited charge and time is generated from each photodetector, as shown in Fig.~\ref{fig:feschematics}. 
For time information, a single-ended to differential amplifier LMH5401 is presented with a gain of 7 V/V, coupled to 50 $\Omega$ of impedance for input and outputs. The differential output signal is sent to the NINO chip discriminator for Time to Digital Conversion (TDC) which is also used as a trigger~\cite{Ugur_2012}.

On the other hand, from Fig.~\ref{fig:feschematics}$b.ii)$, the photocurrent is amplified and integrated through a trans-impedance amplifier OPA858 from Texas Instruments.
In order to process these signals, the TRB3 card is proposed for signal acquisition and processing. Therefore, all collected data can be stored and processed offline.

\clearpage
\section{Simulation studies}
\label{sec:Simulations}

\begin{figure}[!hbt]

\centering\includegraphics[scale=0.6]{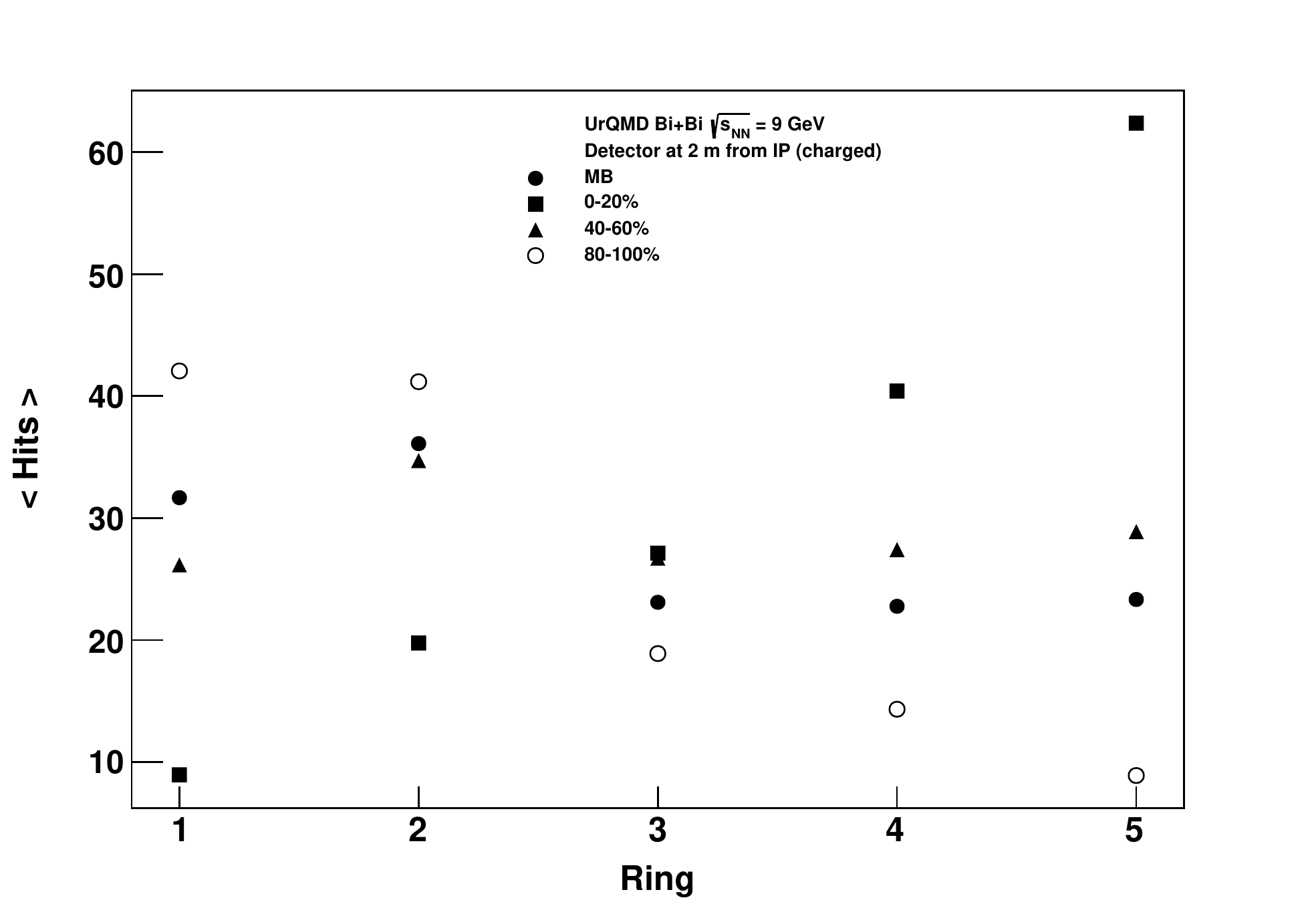}

\caption{Mean number of hits per ring for 1 million MB Bi+Bi collision events at 9 GeV.}
\label{fig:hits-rings-bebe}
\end{figure}

A simulation of the BeBe detector geometry, as described in section~\ref{sec:BeBeDescription}, was performed within the official offline framework of the MPD-NICA experiment, MPDRoot~\cite{MPDROOT}. To evaluate the BeBe detector physics performance for triggering and determination of the resolution of the event plane and centrality of the collision, we performed simulations of 1 million minimum bias (MB) ($b = 0 - 15.9$ fm) events for Bi+Bi collisions at $\sqrt{s_{NN}}=9$ GeV and  Au+Au collisions at $\sqrt{s_{NN}} = 11$ GeV, using the  Ultrarelativistic Quantum Molecular Dynamics (UrQMD) ~\cite{Bass:1998ca,Bleicher:1999xi} which is a microscopic model created to describe hadronic collisions from SIS to AGS, SPS and RHIC energies. UrQMD uses hydrodynamics to describe the evolution of the system from initial stages up to the freeze-out of the produced particles in the collisions and it simulates multiple interactions of newly produced particles and constituents to solve the relativistic Boltzmann equation. Moreover, a simulation of 9,500 MB ($b = 0 - 15.9$ fm) events for Au+Au collisions at $\sqrt{s_{NN}} = 11.5$ GeV with Los Alamos version of the Quark Gluon String model (LAQGSM) ~\cite{Mashnik:2008pt,Mashnik:2016dmf} model was done. LAQGSM model describes those reactions induced by colliding particles as a three-stage process where at the initial stage the primary particles can be re-scattered to produce secondary particles several times prior to the escape or absorption from the nucleus (intranuclear cascade) and then it considers a coalescence model to create high energy particles. The implementation of the BeBe geometry considered only the sensitive material, BC-404, and thus all the analyses were done at the level of simulated hits. As expected, the large density of hits is expected in the two innermost rings of the BeBe detector, see Fig.~\ref{fig:hits-rings-bebe}. With the information given by the number of hits per cell, it is possible to estimate the performance of BeBe detector for centrality and event plane determination. All the simulations considered smearing (\texttt{with}) and no smearing (\texttt{without}) in the vertex simulation.





\subsection{Estimation of BeBe trigger efficiency}
\label{sec:BeBeTrigger}

The main purpose of the BeBe detector is to provide a trigger signal for the identification of Bi+Bi and Au+Au collisions. The raw information of the BeBe trigger signal, generated by either one hit in any of the two BeBe arrays located at opposite sides of the MPD-NICA experiment or two hits in coincidence at a certain time window in both of them, can be used for online luminosity determination of the NICA beam, as it has been demonstrated by the VZERO-ALICE ~\cite{Abbas:2013taa} and LUCID-ATLAS  ~\cite{Manghi:2016jbm} detectors at LHC.

To study the trigger capabilities of BeBe detector two samples of 1 million events were generated using UrQMD Monte Carlo model ~\cite{Bleicher:1999xi,PhysRevC.78.044901} within MPDRoot framework ~\cite{MPDROOT}. We considered two reduced geometries with the closest proposed detector to the MPD interaction point, Mbb~\cite{Kado:2020evi}, FFD, and BeBe detectors. Another geometry configuration considered only Mbb and BeBe. We considered these two configurations to study the effect of FDD material budget on the trigger efficiencies of BeBe detector.
To date, the MPDRoot framework does not include tools to make a detailed simulation of the trigger signals from the MPD-NICA detectors. For this reason, the studies presented in this section assume that BeBe will generate a valid beam-beam trigger signal based on the time of flight of the generated charged particles from the MPD-NICA interaction point reaching the BeBe sensitive cells. Also, flat smearing of 60 cm was considered.

\begin{figure}[!hbt]
\centering
\includegraphics[scale=0.46]{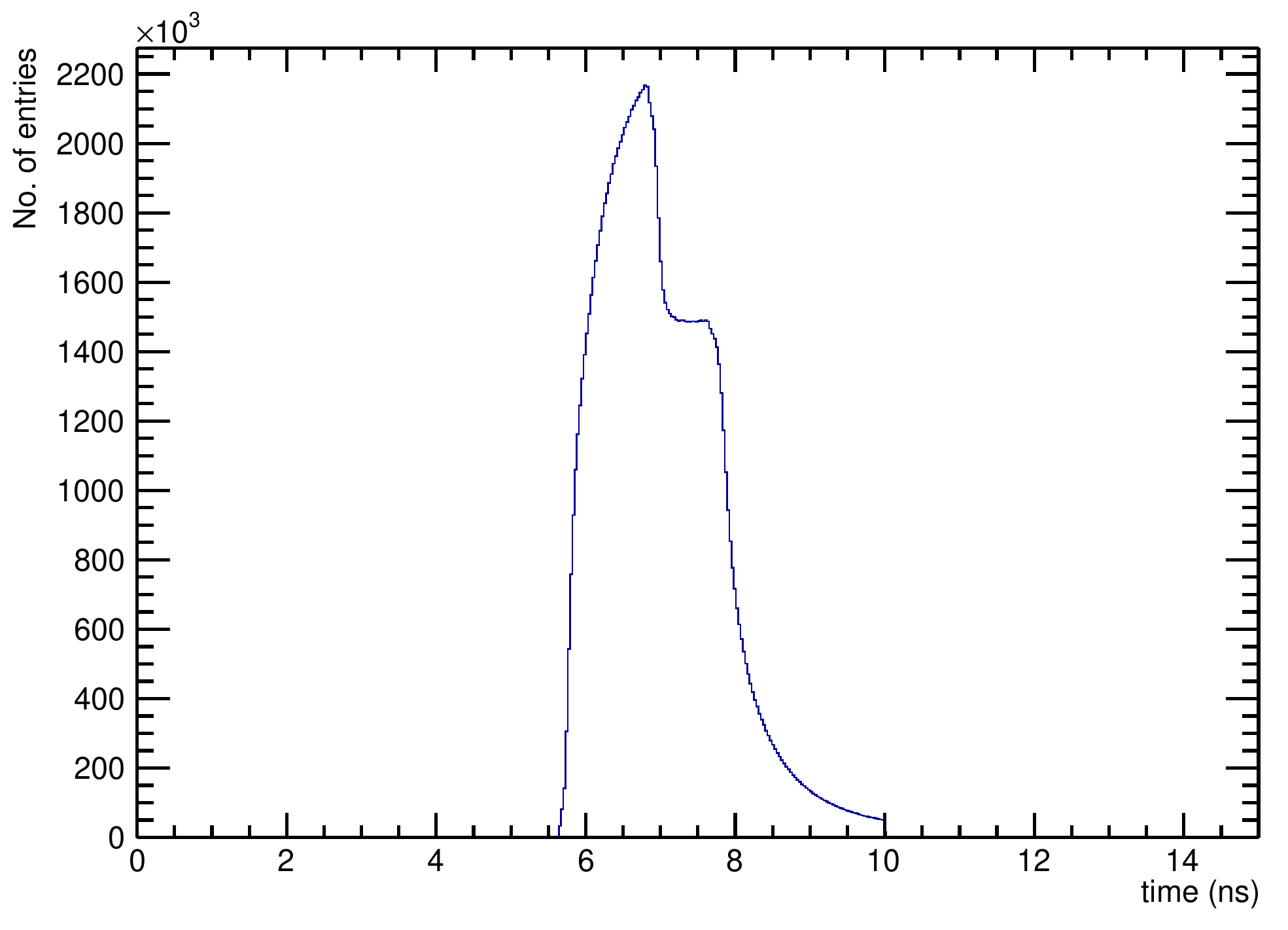}
\caption{Time of flight of the produced charged particles from the interaction point of the collision to BeBe detector on the left side.}
\label{fig:bebetof}
\end{figure}

The elapsed time of flight of the charged particles produced in heavy-ion collisions at NICA, from the interaction point to the BeBe detector cells of any of its two arrays, on average is of the order of 7 ns, see Fig.~\ref{fig:bebetof}. To simulate the BeBe trigger signals for heavy-ion collisions at NICA the time of flight information of the BeBe simulated hits was used. 

In Fig.~\ref{fig:bebetof}, we defined a time window of $\Delta \tau = 7 ns +/- 3 ns$ to the time of flight distribution for each matrix to simulate the following  BeBe trigger flags:

\begin{itemize}
    \item \texttt{BBR}: if the $Z$ coordinate of the BeBe hit is positive and the time of flight of the first BeBe hit is within the time window defined by $\Delta \tau$.
    \item \texttt{BBL}: if the $Z$ coordinate of the BeBe hit is negative and the time of flight of the first BeBe hit is within the time window defined by $\Delta \tau$.
    \item \texttt{BBR AND BBL}: logical \texttt{AND} of the coincidence of \texttt{BBR} and \texttt{BBL}.
    \item \texttt{BBR OR BBL}: logical \texttt{OR} of \texttt{BBR} and \texttt{BBL}.
\end{itemize}

For p+p collisions at $\sqrt{s}=9$~GeV and $\sqrt{s}=11$~GeV the trigger efficiency given either by \texttt{BBR} or \texttt{BBL} is of the order of 58\%  if a vertex smearing is assumed (\texttt{with}) in the simulation. Our results suggest that both trigger efficiencies increase up to 73\% when no smearing on the vertex simulation is considered. 
Thus, the \texttt{BBL} and \texttt{BBR} trigger efficiencies will have a strong dependence on the vertex smearing, which is directly related to the NICA beam quality. The  \texttt{BBR AND BBL} and \texttt{BBR OR BBL} trigger efficiencies are 20.26\% and 95.6\% respectively for p+p collisions at $\sqrt{s}=9$ GeV. For heavy-ion collisions, the trigger efficiencies are larger than 90\% in all the assumed configurations. 

The BeBe detector can contribute to construct a minimum bias trigger for MPD experiment with efficiencies of at least 95\% using the trigger condition \texttt{BBR OR BBL}. The BeBe trigger \texttt{BBR AND BBL} can be used for luminosity studies to determine the visible cross-section estimating the number of events with at least one hit on both sides of BeBe detector. The efficiency of \texttt{BBR AND BBL} trigger varies from 20.26\% (vertex smearing) to 50.12\% (no vertex smearing). Depending on the NICA beam profile the efficiency of BeBe detector can be compromised for luminosity studies of the NICA beam. More details about luminosity studies at LHC with forward detectors can be found in ~\cite{ALICE-PUBLIC-2016-002}.

The estimation of the trigger efficiencies of BeBe detector seems to be independent of the Monte Carlo generator used. In this case, we estimated the BeBe trigger efficiencies with UrQMD ~\cite{Bass:1998ca,Bleicher:1999xi} and LAQGSM models ~\cite{Mashnik:2008pt}, see table ~\ref{Tab:LAQGSMEff}. Moreover, as the FDD detector is located in front of BeBe detector from the IP of MPD it induces a production of secondary particles that reduces the BeBe trigger efficiencies: $\sim$ 4\% for \texttt{BBR AND BBL} in p+p at 9 GeV, $\sim$ 10\% for \texttt{BBR AND BBL} in Bi+Bi at 9 GeV and $\sim$2\% for \texttt{BBR OR BBL} in Bi+Bi at 9 GeV when a vertex smearing is considered, see tables~\ref{Tab:Case1FfdOn} and ~\ref{Tab:Case2FfdOff}.


\begin{table}[]
\centering
{
\begin{tabular}{|l|l|l|l|l|l|}
\hline
\multicolumn{1}{|c|}{Process}  & BBR       & BBL      & BBRandBBL & BBRorBBL & Vertex smearing \\ \hline

p+p, 9 GeV            & 58.063\%  & 57.86\%  & 20.26\%   & 95.66\%  & Yes \\ \hline
p+p, 9 GeV           & 72.85\%   & 72.79\%  & 50.12\%   & 95.52\% & No  \\ \hline
p+p, 11 GeV            & 59.84\%   & 59.87\%  & 23.41\%   & 95.52\%  & Yes \\ \hline
p+p, 11 GeV           & 74.31\%   & 74.42\%  & 52.7\%    & 96.03\% & No \\ \hline
Bi+Bi, 9 GeV           & 94.07\%   & 94.07\%  & 89.88\%   & 98.26\% & Yes \\ \hline
Bi+Bi, 9 GeV          & 100\%     & 100\%    & 100\%     & 100\% & No    \\ \hline
Au+Au, 11 GeV          & 100\%     & 100\%    & 100\%     & 100\% & Yes   \\ \hline
Au+Au, 11 GeV          & 100\%     & 100\%    & 100\%     & 100\% & No   \\ \hline
\end{tabular}
\caption{Trigger efficiencies of BeBe. Simulated detectors: Mbb, FFD, and BeBe.}
\label{Tab:Case1FfdOn}
}

\end{table}

\begin{table}[]
\centering
{
\begin{tabular}{|l|l|l|l|l|l|}
\hline
\multicolumn{1}{|c|}{Process}  & BBR       & BBL      & BBRandBBL & BBRorBBL & Vertex smearing \\ \hline
pp@9GeV            & 56.07\%  & 57.86\%  & 16.79\%   & 95.17\% & Yes \\ \hline
pp@9GeV           & 71.99\%   & 72.05\%  & 49.01\%   & 95.03\% & No \\ \hline
pp@11GeV           & 57.66\%   & 57.46\%  & 19.26\%   & 95.85\% & Yes \\ \hline
pp@11GeV           & 73.35\%   & 73.43\%  & 51.25\%    & 95.53\% & No \\ \hline
BiBi@9GeV          & 100\%   & 100\%  & 100\%   & 100\% & Yes \\ \hline
BiBi@9GeV         & 100\%     & 100\%    & 100\%     & 100\% & No    \\ \hline
AuAu@11GeV          & 100\%     & 100\%    & 100\%     & 100\% & Yes   \\ \hline
AuAu@11GeV          & 100\%     & 100\%    & 100\%     & 100\%  & No  \\ \hline
\end{tabular}
\caption{Trigger efficiencies of BeBe. Simulated detectors: Mbb and BeBe.}
\label{Tab:Case2FfdOff}
}
\end{table}

\begin{table}[]
\centering
{
\begin{tabular}{|l|c|c|c|c|}
\hline
\multicolumn{1}{|c|}{Process}  & BBR       & BBL      & BBRandBBL & BBRorBBL \\ \hline
AuAu@11.5GeV        & 97.7\%     & 97.6\%    & 95.4\%     & 99.9\%    \\ \hline
\end{tabular}
\caption{Trigger efficiencies of BeBe, using LAQGSM. Simulated detectors: Mbb, BeBe, FHCAl, and FFD.}
\label{Tab:LAQGSMEff}
}
\end{table}


\subsection{Centrality determination}
\label{sec:BeBecentrality}

Centrality is a key variable for characterizing the geometric properties of the heavy-ion collisions. Many experimental techniques devoted to the study of the nuclear matter created in ultrarelativistic heavy-ion collisions depend on the trigger and reconstructed information provided by forward detectors such as BeBe. This has been demonstrated in recent years by ALICE-LHC experiment~\cite{ALICEcentrality}. 

\begin{figure*}[!hbt]

\centering\includegraphics[scale=0.4]{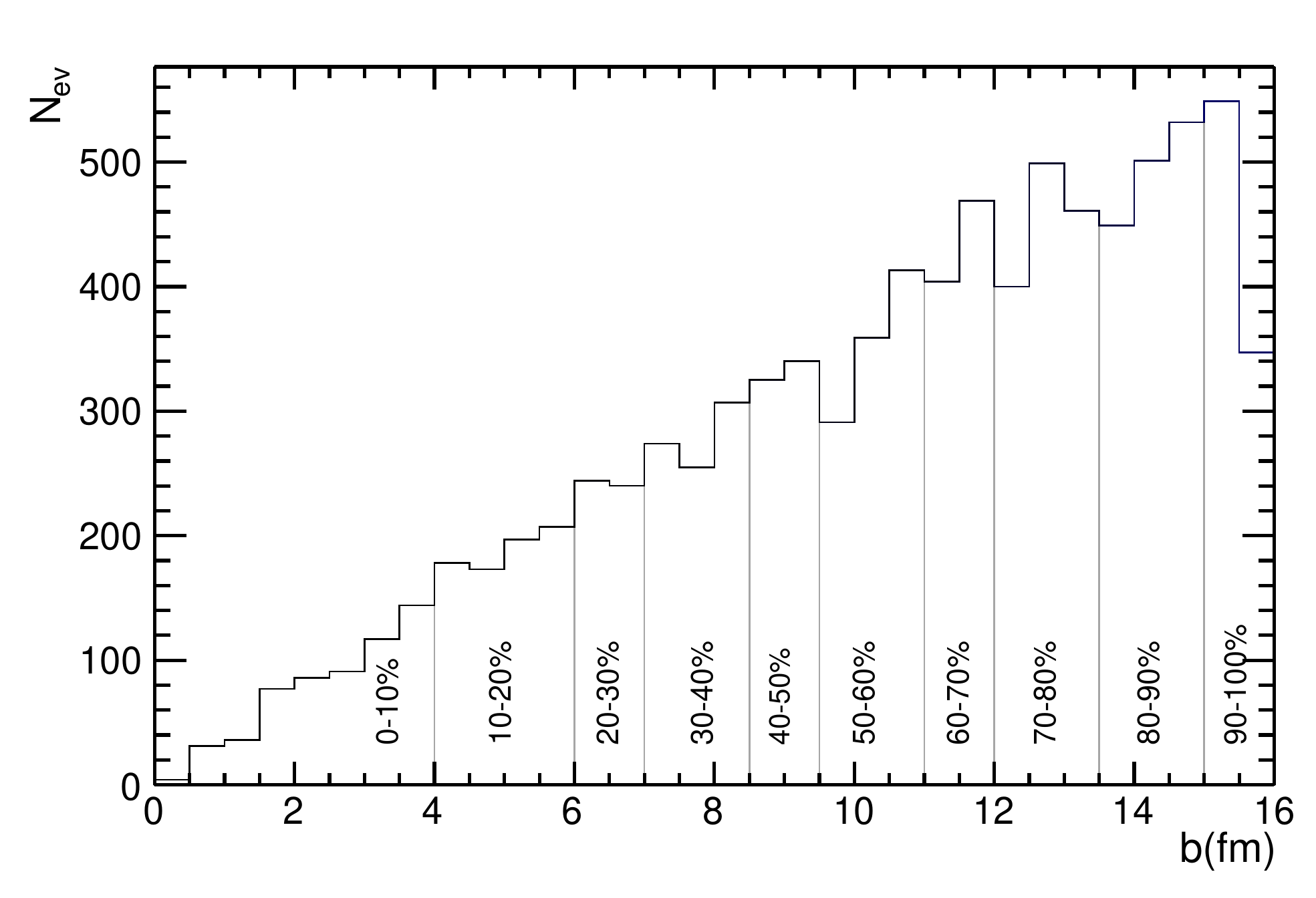}\includegraphics[scale=0.4]{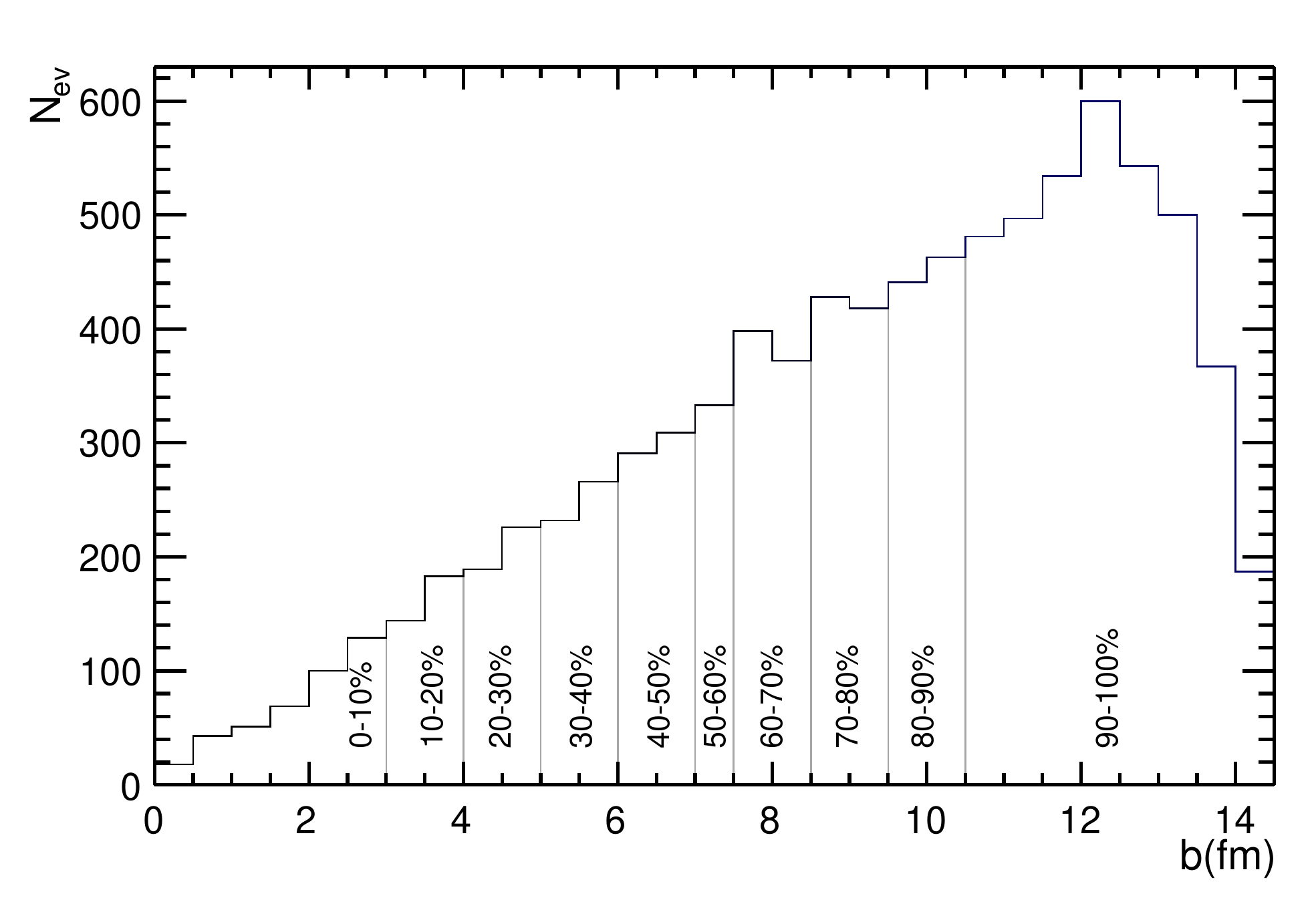}

\centering\includegraphics[scale=0.4]{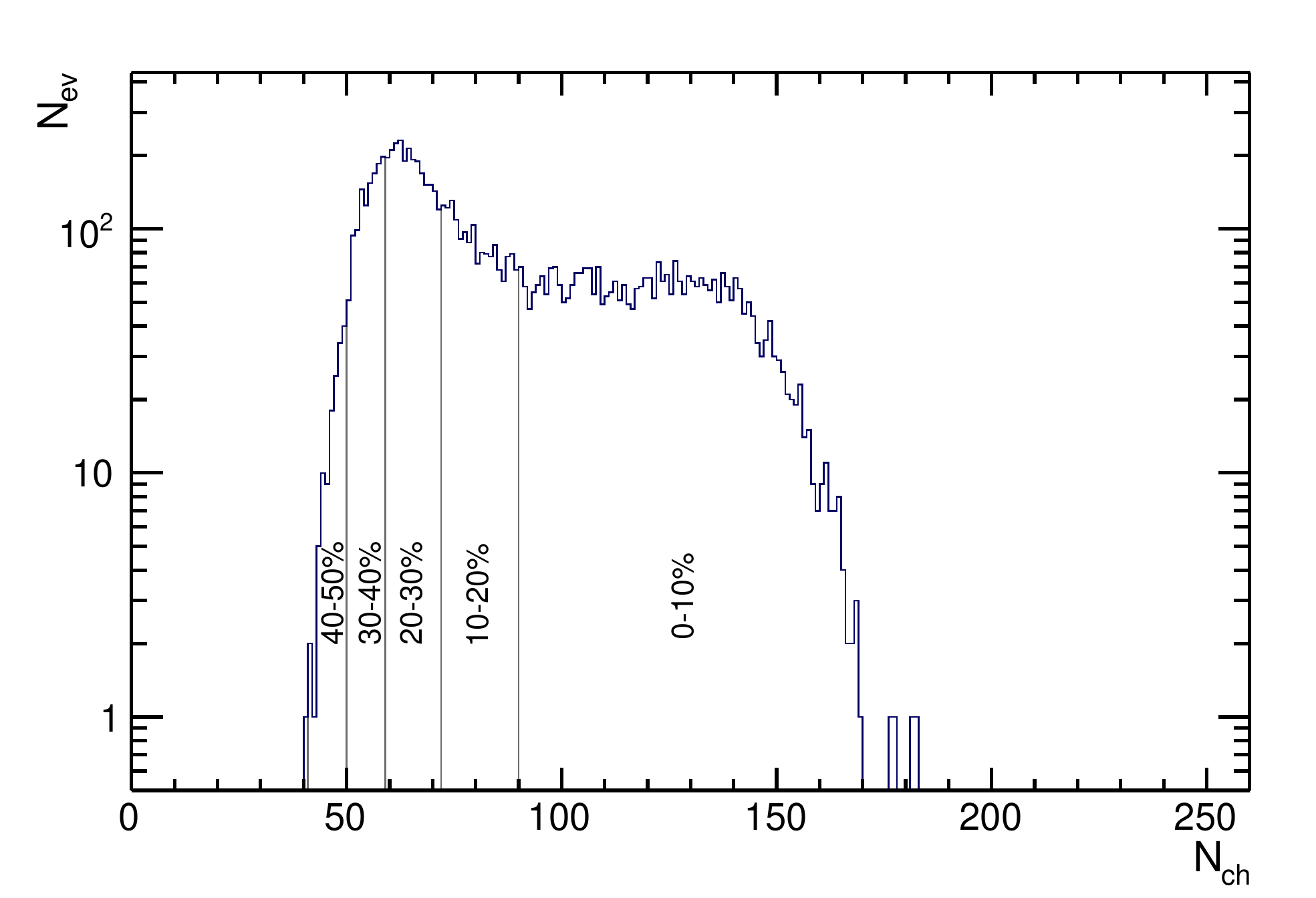}\includegraphics[scale=0.4]{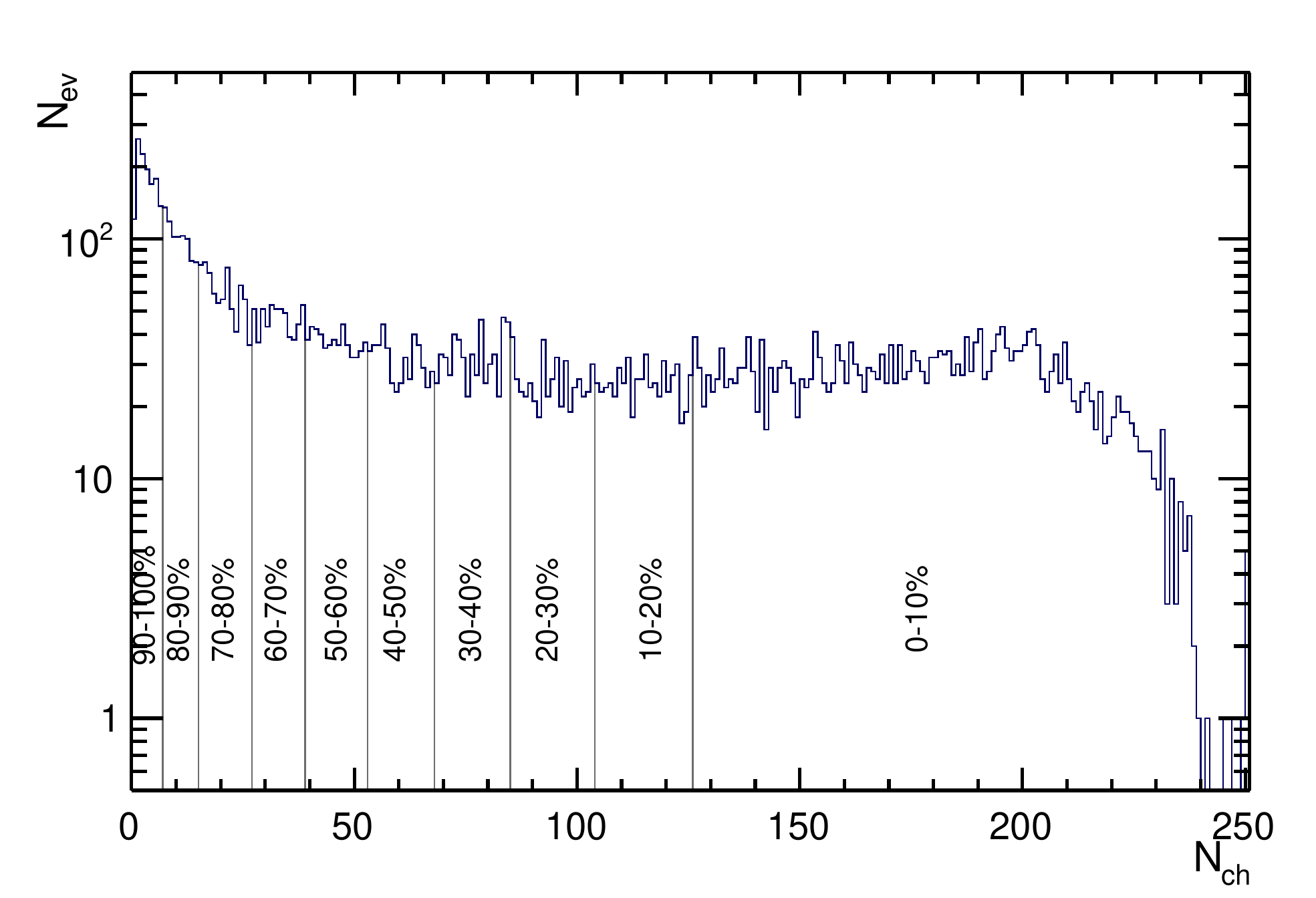}

\caption{Centrality classes, impact parameter ranges (top), and the number of hits (bottom) for 9,500 MB Au+Au at 11 GeV with UrQMD (left) and Au+Au at 11.5 GeV with LAQGSM (right), using all rings of BeBe.}
 \label{MultiplicityClassesall}
\end{figure*}

\begin{figure*}[!hbt]


\centering\includegraphics[scale=0.4]{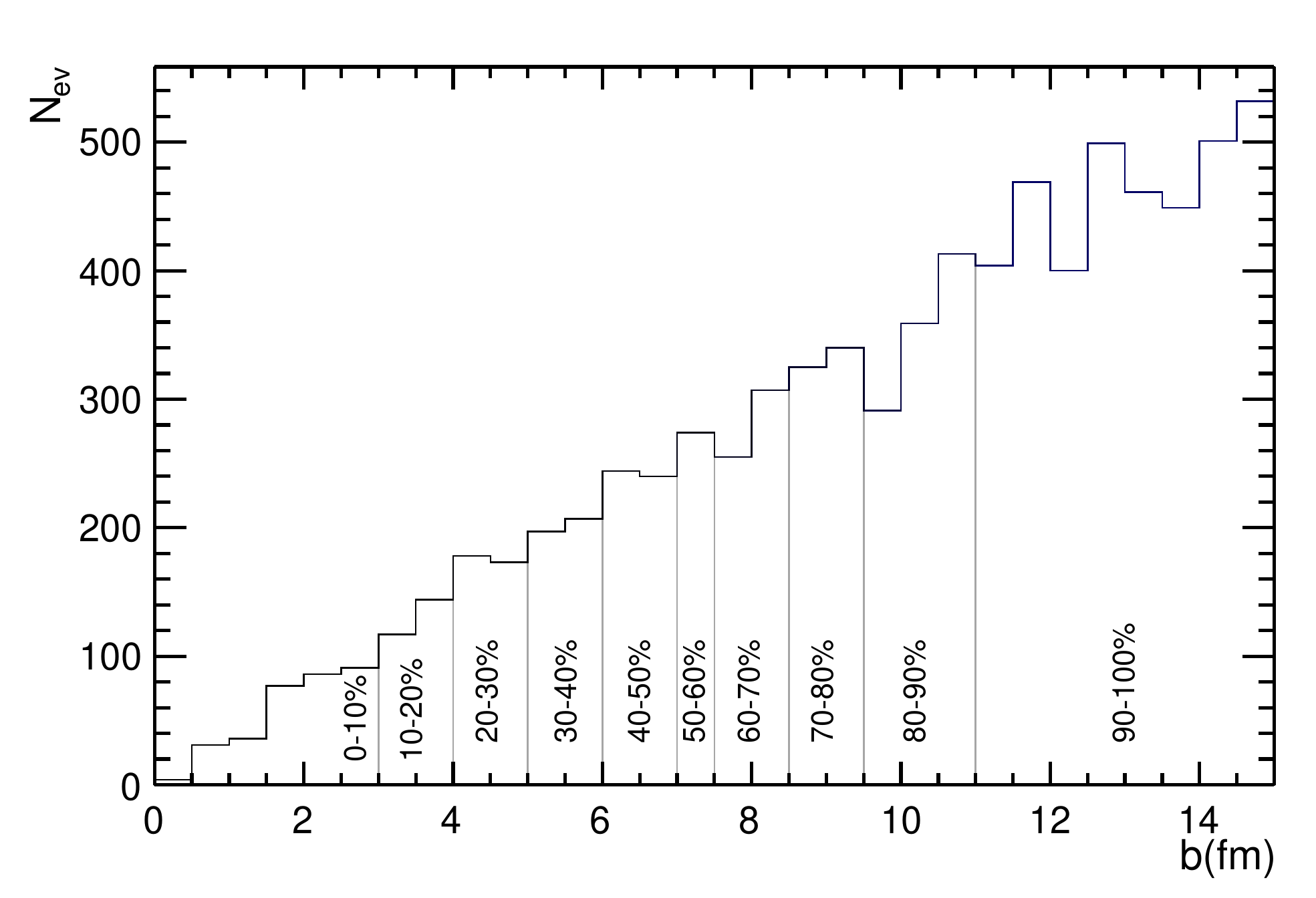}\includegraphics[scale=0.4]{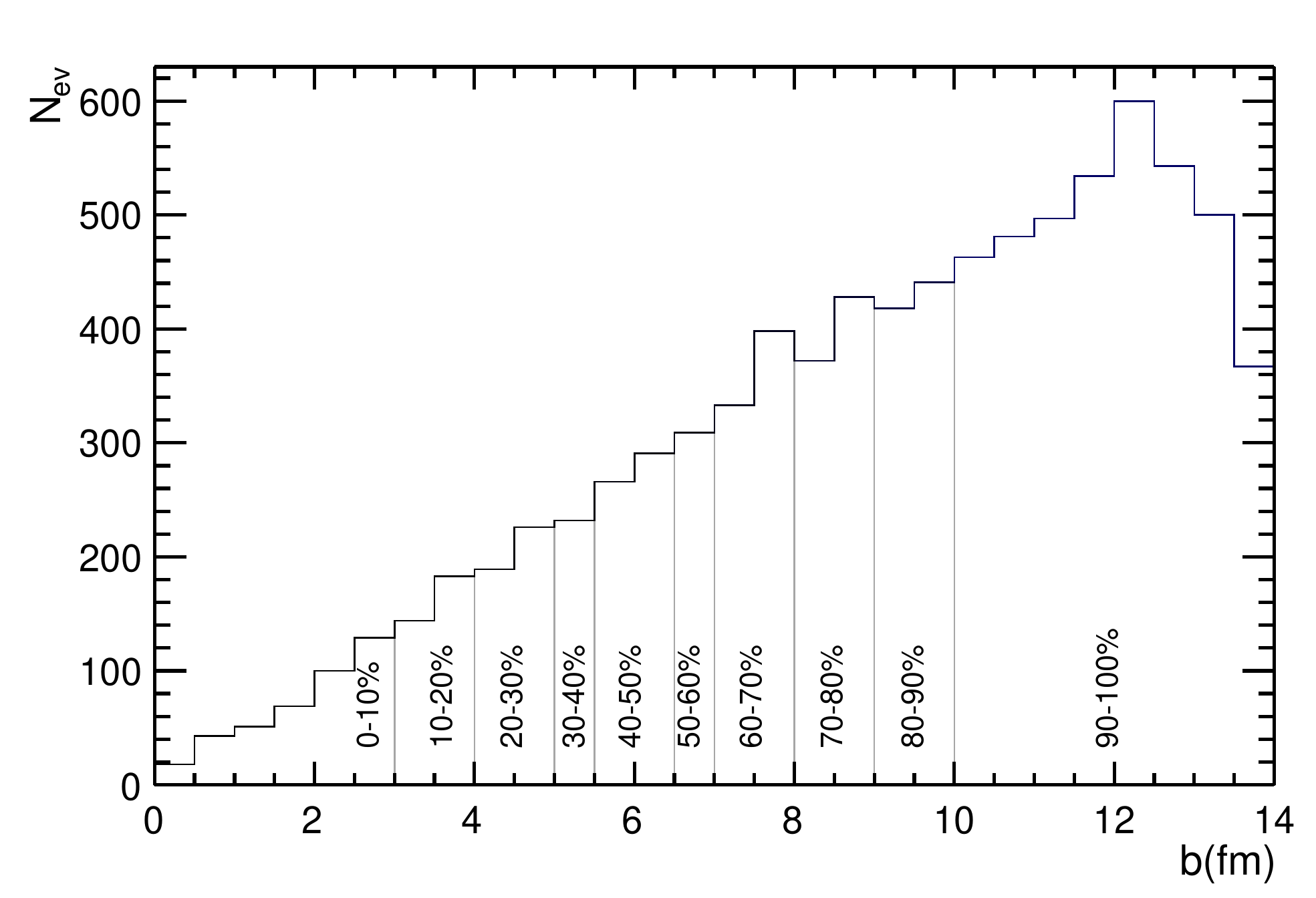}


\centering\includegraphics[scale=0.4]{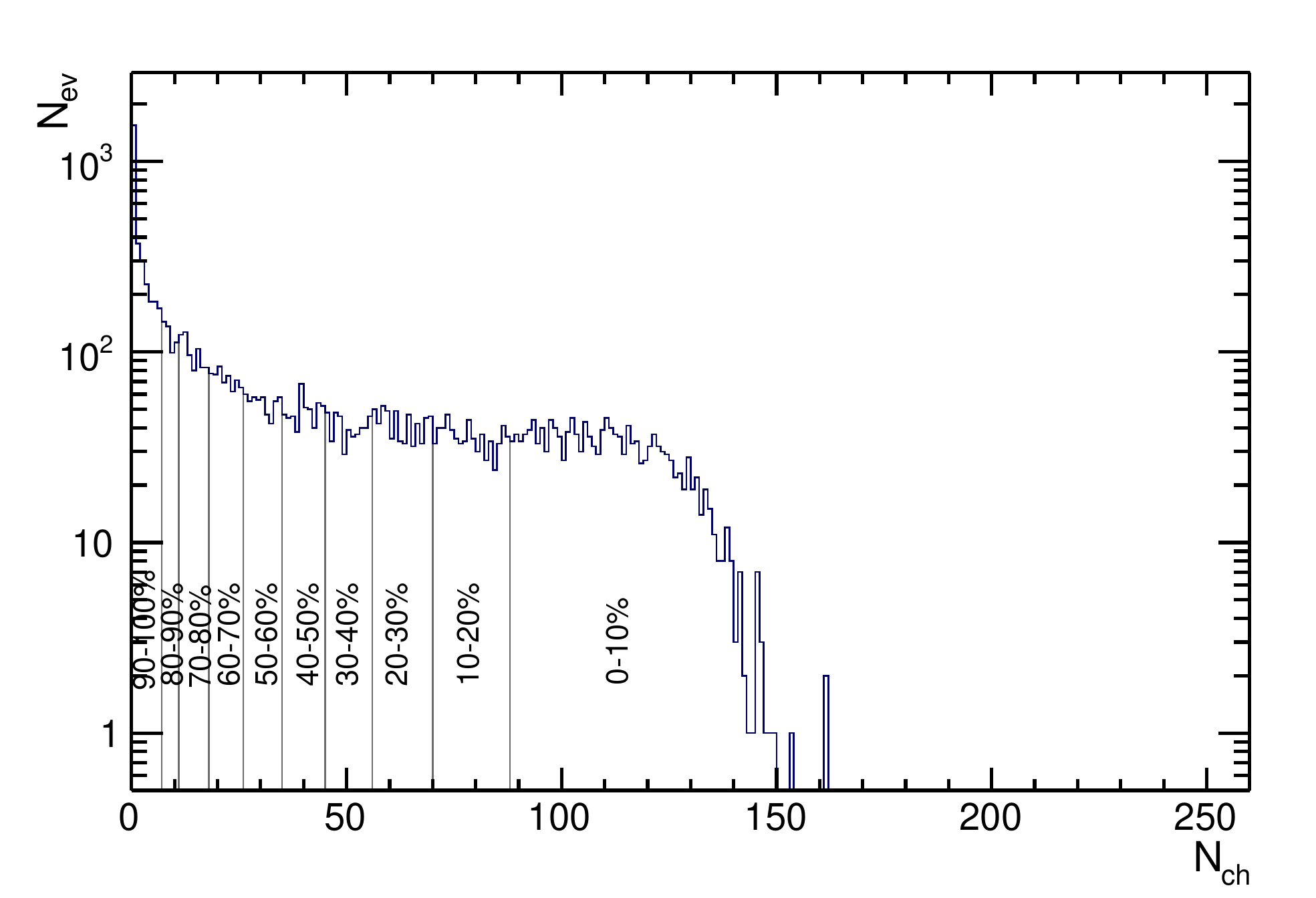}\includegraphics[scale=0.4]{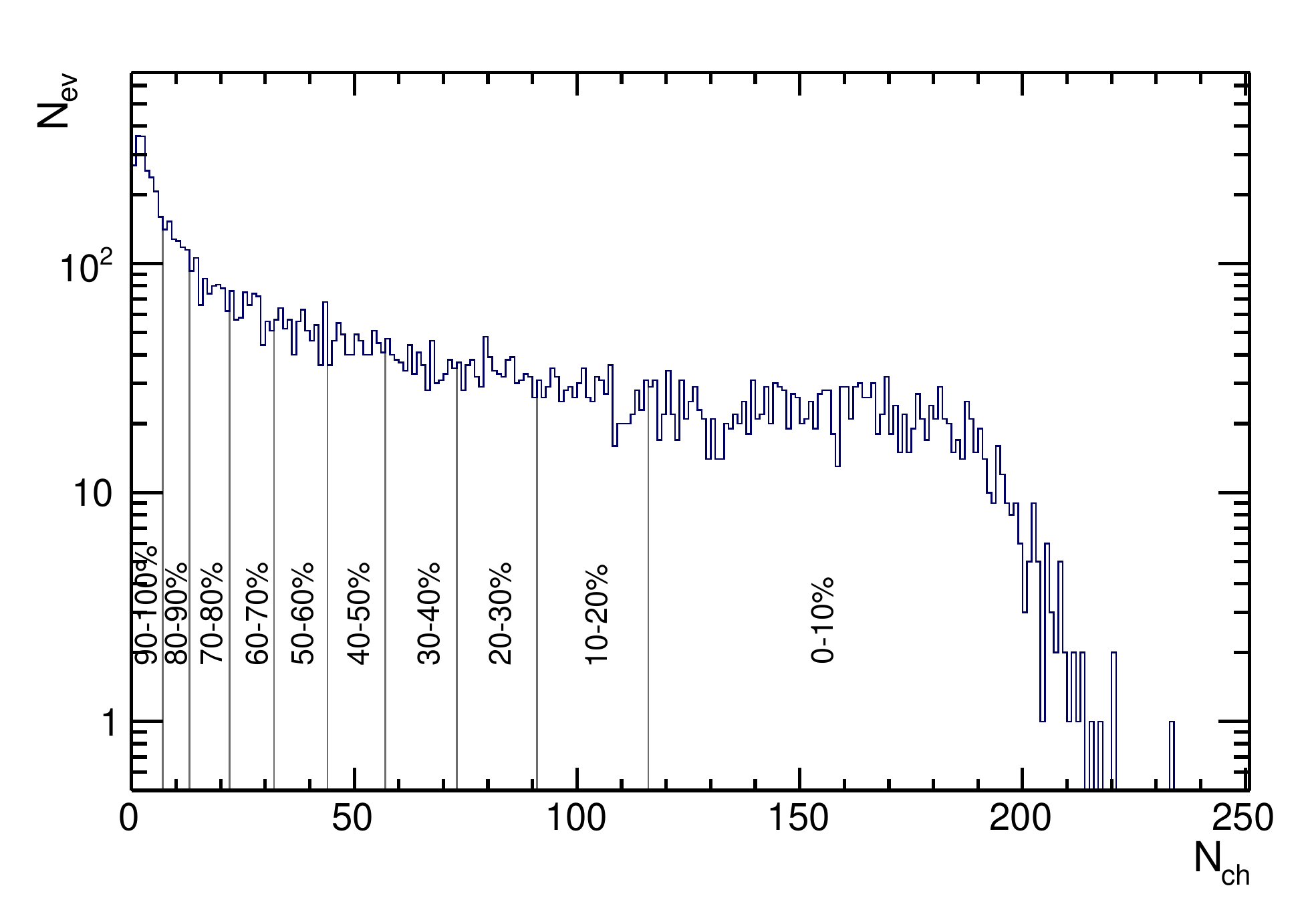}

\caption{Centrality classes, impact parameter ranges (top), and the number of hits (bottom) for 9,500 MB Au+Au at 11 GeV with UrQMD (left) and Au+Au at 11.5 GeV with LAQGSM (right), using rings 3-5 rings of BeBe.}
 \label{MultiplicityClasses345}
\end{figure*}

To estimate the BeBe detector capabilities in centrality determination of the heavy-ion collisions at NICA energies, an UrQMD and LAQGSM simulations of 9,500 Minimum Bias Au+Au collision events at $\sqrt{s_{NN}}=$ 11 GeV were generated within the MPDroot framework. As a first step, we computed the number of charged particles (denoted by $N_{ch}$) reaching the BeBe detector cells, as a function of the simulated impact parameter. In the case of plastic scintillator detectors, it has been shown by VZERO-ALICE ~\cite{Abbas:2013taa} at LHC that the shape of the number of hits in the detector can be described in terms of the Glauber model ~\cite{Miller:2007ri}. With the proposed geometry for BeBe detector, we observe that it is not a good option to employ all the five rings of BeBe detector, UrQMD prediction. This behavior is in contrast with the prediction given by LAQGSM model where the BeBe hits distribution exhibits a nice curve that can be adjusted by a Glauber-like function, see Fig.~\ref{MultiplicityClassesall}. For UrQMD, this situation improves if we only take into account the hit multiplicity of the three outer rings of BeBe, Fig.~\ref{MultiplicityClasses345}. In tables~\ref{tabla:Clasesbmul-urqmdbb} and ~\ref{tabla:Clasesbmul-bbLAQGSM} the minimum and maximum values of hit multiplicities in BeBe detector are given. These values were obtained inspired by the method discussed in~\cite{ALICEcentrality}.

\begin{table*}[!hbt]
\centering
\begin{tabular}{ccccc}
\hline
\multicolumn{5}{c}{Au+Au@11GeV, UrQMD all BeBe rings}\\
\hline\hline
Class \% & $b$ min (fm) & $b$ max  (fm) & $N_{ch}$ max & $N_{ch}$ min \\ 
 \small 0-10 & 0&3.9105             &  180  & 90 \\                
 \small 10-20 & 3.9105&5.5195           &  90  & 72 \\                 
 \small 20-30 & 5.5195&6.8495           &  72  & 59 \\                 
 \small 30-40 & 6.8495&8.1495             &  59  & 50 \\                 
 \small 40-50 & 8.1495&9.3295            &  50  & 41 \\                 
 \small 50-60 & 9.3295&10.6495           &  41  & 32\\                  
 \small 60-70 & 10.6495&11.8805          &  32  & 23 \\                 
 \small 70-80 & 11.8805&13.1805          &  23  & 14\\                  
 \small 80-90 & 13.1805&14.5605         &  14  & 7 \\                  
 \small 90-100 & 14.5605&15.8105         &  7  & 0 \\                   
\hline
\multicolumn{5}{c}{Au+Au@11GeV, UrQMD, 3-5 BeBe rings}\\
\hline\hline
Class \% & $b$ min (fm) & $b$ max  (fm) & $N_{ch}$ max & $N_{ch}$ min \\ 
 \small 0-10  & 0      & 2.7795         &  160  & 88 \\                  
 \small 10-20 & 2.7795 & 3.9605          &  88  & 70 \\                  
 \small 20-30 & 3.9605 & 4.8695          &  70 & 56 \\                  
 \small 30-40 & 4.8695 & 5.7395          &  56  & 45 \\                  
 \small 40-50 & 5.7395 & 6.5505          &  45  & 35 \\                  
 \small 50-60 & 6.5505 & 7.4005          &  35  & 26\\                   
 \small 60-70 & 7.4005 & 8.3005          &  26  & 18 \\                  
 \small 70-80 & 8.3005 & 9.3005          &  18  & 12\\                   
 \small 80-90 & 9.3005 & 10.6895         &  12  & 7 \\                   
 \small 90-100 & 10.6895 & 14.9205         &  7  & 0 \\                  
\hline
\end{tabular}

\caption{Centrality classes and impact parameter ranges and number of charged particles in 9,500 events of Au+Au at 11 GeV UrQMD using all rings (top) and 3-5 rings (bottom).}
\label{tabla:Clasesbmul-urqmdbb}
\end{table*}

\begin{table*}[!hbt]
\centering
\begin{tabular}{ccccc}
\hline
\multicolumn{5}{c}{Au+Au@11.5GeV, LAQGSM all BeBe rings}\\
\hline\hline
Class \% & $b$ min (fm) & $b$ max  (fm) & $N_{ch}$ max & $N_{ch}$ min \\ 
 \small 0-10  & 0      & 2.7795             &  265  & 140 \\                
 \small 10-20 & 2.7795 &  3.9795            &  140  & 114 \\                 
 \small 20-30 &  3.9795 & 4.9405            &  114  & 92 \\                 
 \small 30-40 & 4.9405 & 5.8135             &  92  & 73 \\                 
 \small 40-50 & 5.8135 & 6.6355             &  73  & 57 \\                 
 \small 50-60 & 6.6355 & 7.3945             &  57  & 42\\                  
 \small 60-70 & 7.3945 & 8.1735             &  42  & 28 \\                 
 \small 70-80 & 8.1735 & 9.0905             &  28  & 16\\                  
 \small 80-90 & 9.0905& 10.2775             &  16  & 7 \\                   
 \small 90-100 & 10.2775 & 14.0195          &  7  & 0 \\                   
\hline
\multicolumn{5}{c}{Au+Au@11.5GeV, LAQGSM 3-5 BeBe rings}\\
\hline\hline
Class \% & $b$ min (fm) & $b$ max  (fm) & $N_{ch}$ max & $N_{ch}$ min \\ 
 \small 0-10  & 0      & 2.5255         &  250  & 120 \\                  
 \small 10-20 & 2.5255 & 3.6115          &  120  & 94 \\                  
 \small 20-30 & 3.6115 & 4.5005          &  94  & 74 \\                  
 \small 30-40 & 4.5005 & 5.2915          &  74  & 58 \\                  
 \small 40-50 & 5.2915 & 6.1165          &  58  & 45 \\                  
 \small 50-60 & 6.1165 & 6.9195          &  45  & 33\\                   
 \small 60-70 & 6.9195 & 7.7325          &  33  & 23 \\                  
 \small 70-80 & 7.7325 & 8.7395          &  23  & 13\\                   
 \small 80-90 & 8.7395 & 9.9895         &  13  & 7 \\                   
 \small 90-100 & 9.9895 & 13.9445         &  7  & 0 \\                  
\hline
\end{tabular}
\caption{Centrality classes, impact parameter ranges, and the number of charged particles in 9,500 events of Au+Au at 11.5 GeV LAQGSM  using all rings (top) and 3-5 rings (bottom).}
\label{tabla:Clasesbmul-bbLAQGSM}
\end{table*}





\subsection{Centrality resolution}
\label{sec:BeBeResolcentrality}

\begin{figure*}[!hbt]
\centering\includegraphics[scale=0.4]{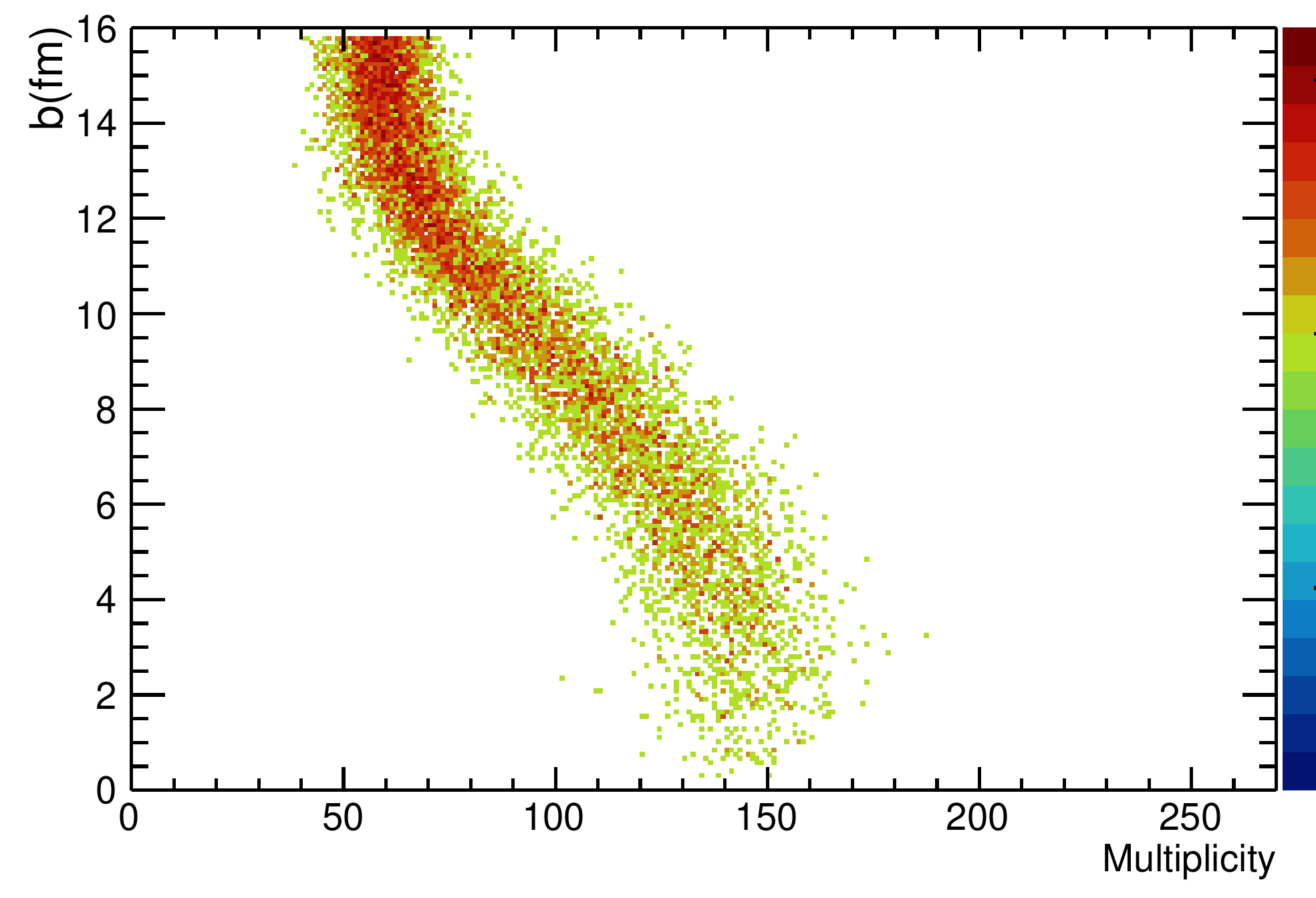}\includegraphics[scale=0.4]{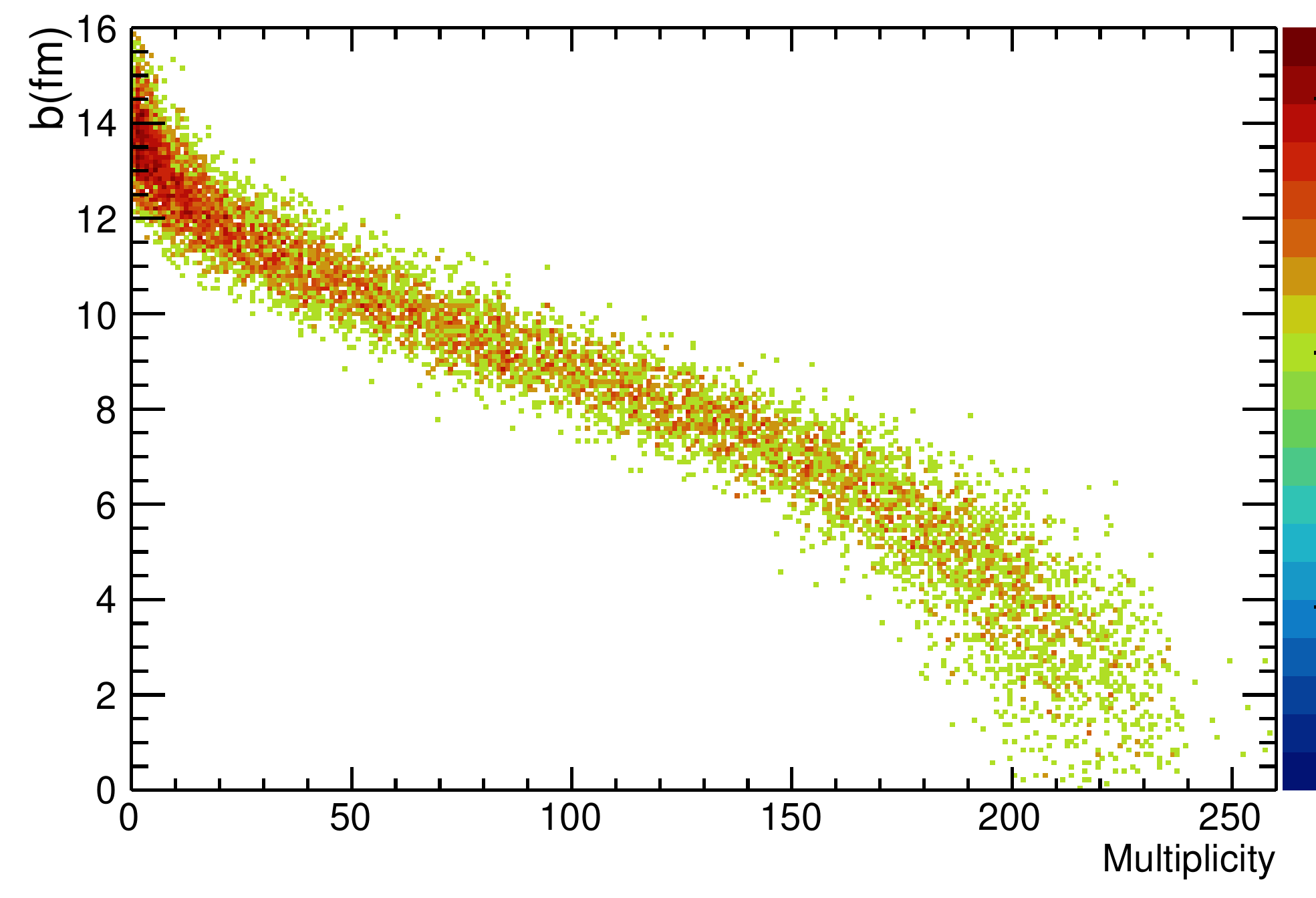}

\centering\includegraphics[scale=0.4]{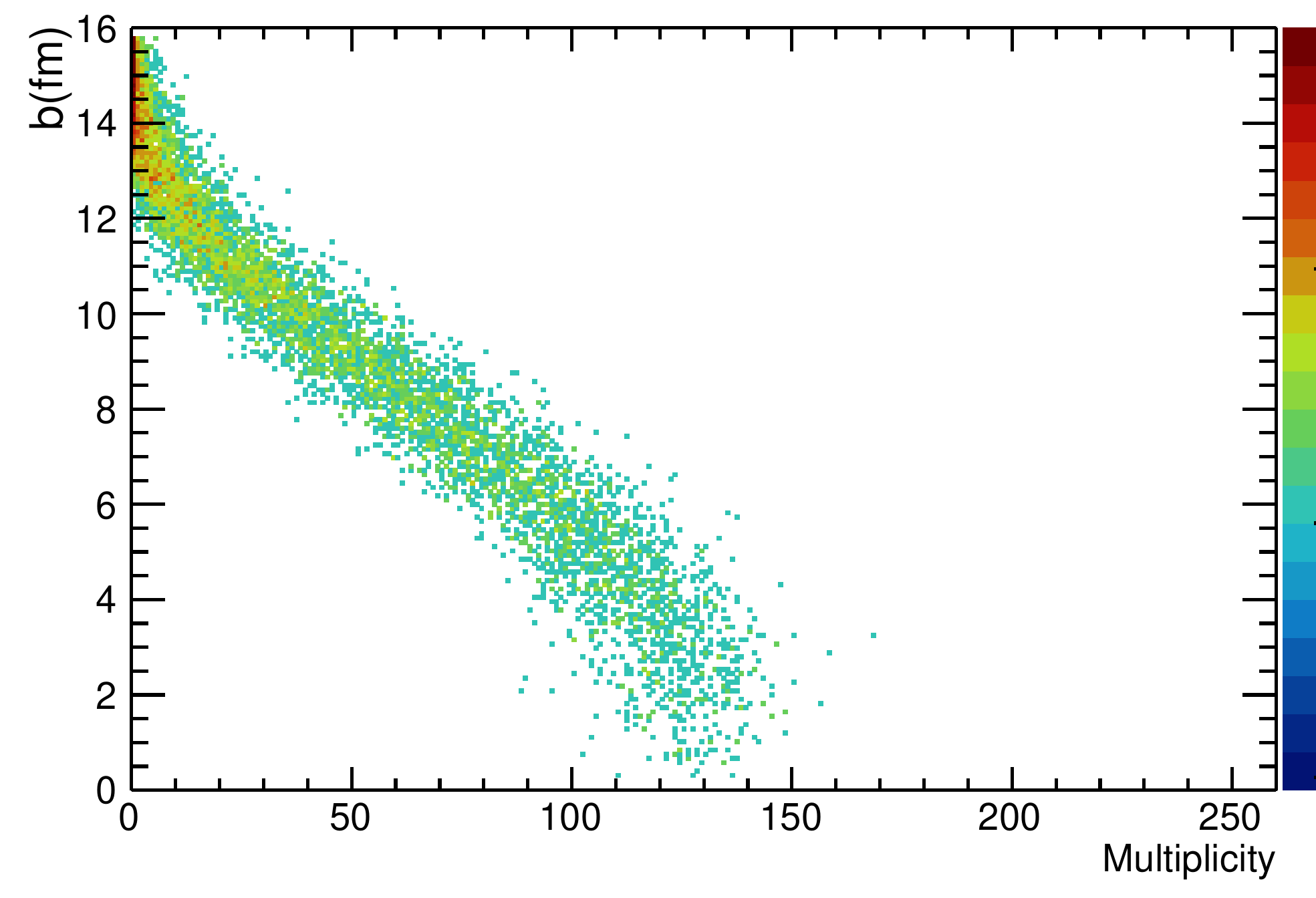}\includegraphics[scale=0.4]{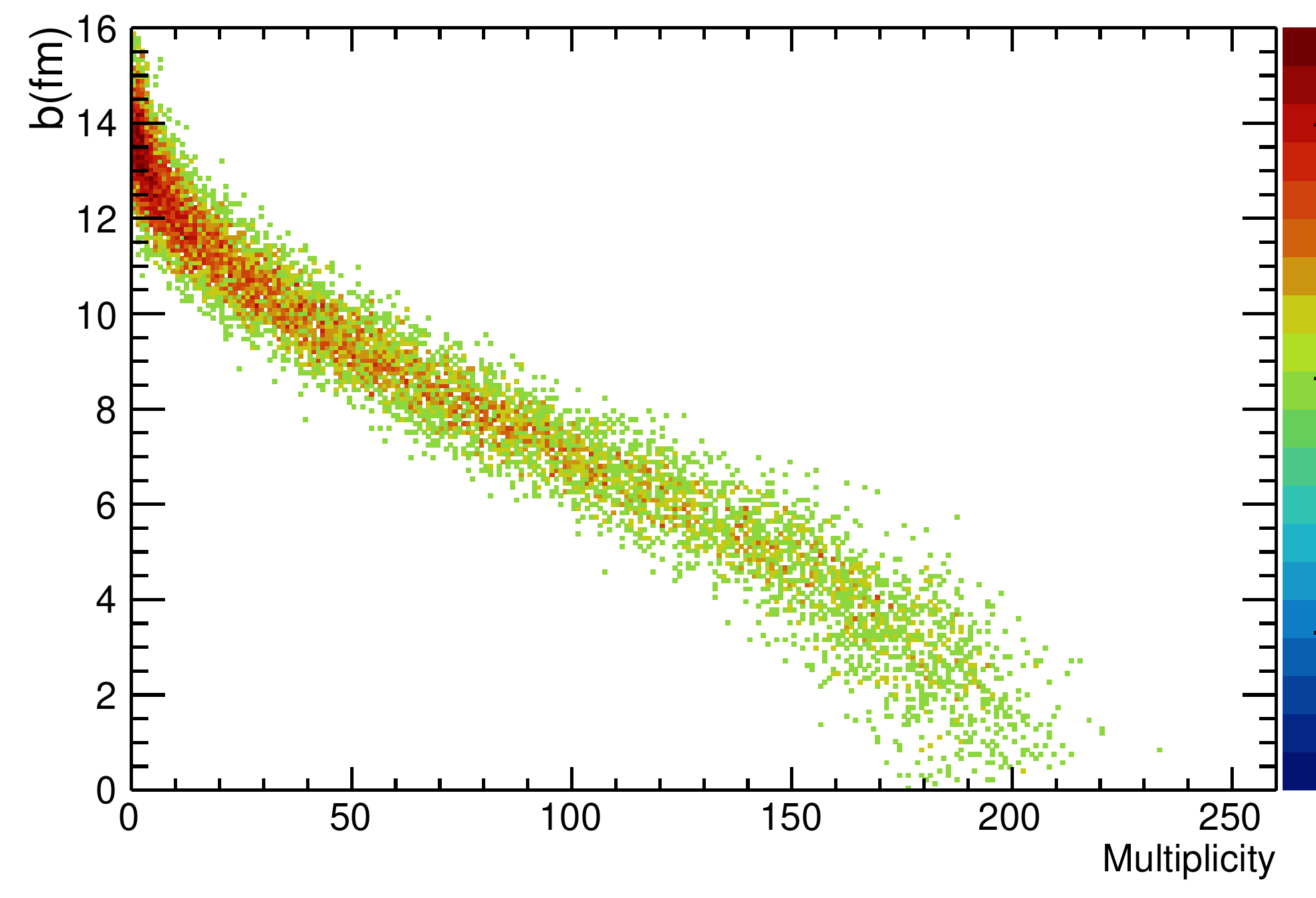}

\caption{Correlation between multiplicity and impact parameter for all BeBe rings (top) and 3-5 BeBe rings (bottom) for 9,500 MB Au+Au at 11 GeV with UrQMD (left) and Au+Au at 11.5 GeV with LAQGSM (right).}
 \label{CorrelationNBall345rings}
\end{figure*}

To compute the centrality resolution given by BeBe detector, we correlated the generated impact parameter with the hit multiplicity in BeBe and looked for the best curve behavior, a linear correlation. As can be seen in Fig.~\ref{CorrelationNBall345rings}, such linear correlation is predicted by LAQGSM model independently of the number of BeBe rings used.


\begin{figure}[!hbt]
\centering

\includegraphics[scale=0.54]{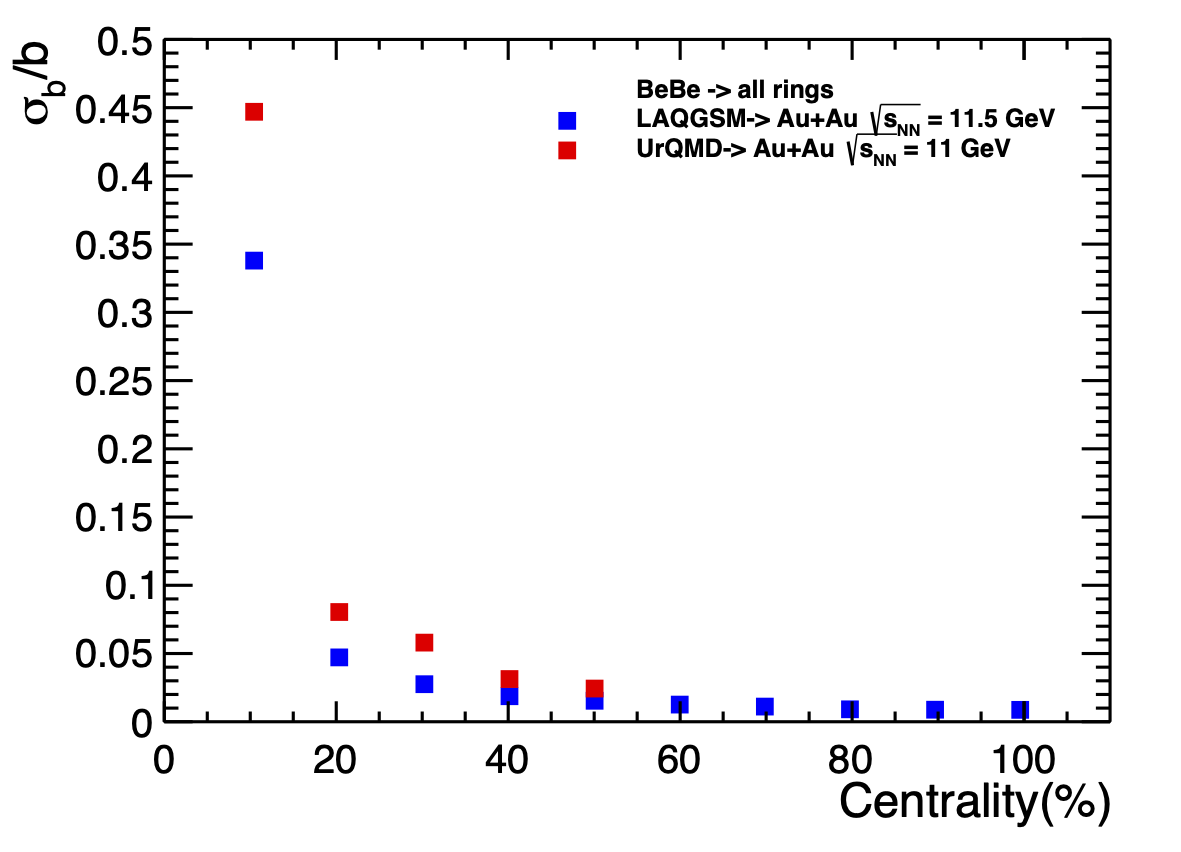}
\caption{Centrality resolution for 9,500 MB events for Au+Au@11 GeV with UrQMD and 9,500 MB events for Au+Au@11.5 GeV with LAQGSM, using in both cases all BeBe rings.}
\label{fig:BBGeometry}
\end{figure}

The prediction given by UrQMD suggests employing only the three outer rings of BeBe detector. With the computed values shown in tables~\ref{tabla:Clasesbmul-urqmdbb} and ~\ref{tabla:Clasesbmul-bbLAQGSM} we estimated a mean value of the centrality using the number of hits in BeBe detector. This value can be compared with the truth value of the centrality given by the generated impact parameter. Event by event, we computed the difference between the centrality given by the number of hits in the BeBe detector ($cent_{BeBe}$) with respect to the generated centrality ($cent_{MC}$), $cent=cent_{MC}-cent_{BeBe}$. The width of a Gaussian fit of $cent$ distribution will give us the centrality resolution of BeBe detector with respect to the centrality of the collision. This method is based on an ideal performance of the proposed detector and we do not take into account detector inefficiencies from the whole data acquisition chain, underestimation of secondaries due to material budget, or reconstruction effects.  In Fig.~\ref{fig:BBGeometry}, the centrality resolution of BeBe detector for UrQMD and LAQGSM models is shown. Using the hit multiplicity of all the BeBe detector rings, UrQMD model predicts a centrality resolution of 45\% and LAQGSM model prediction is 34 \% for central collision. As the percentage of the centrality increases, the centrality resolution given by BeBe improves up to 5\% for both models. 

\begin{figure}[!hbt]
\centering

\includegraphics[scale=0.45]{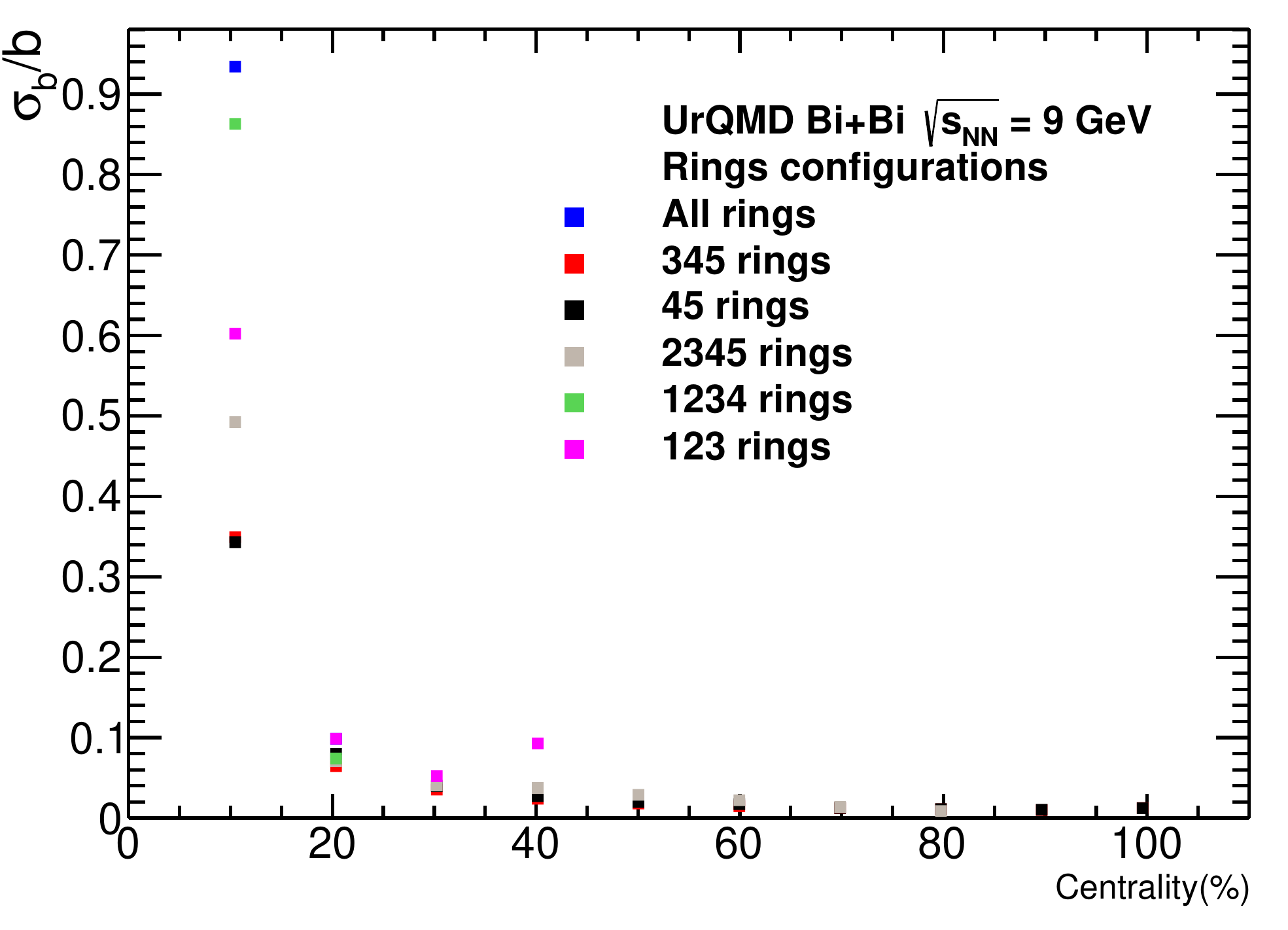}
\caption{Centrality resolution for several BeBe rings configurations.}
\label{fig:BBCentRingsConfiguration}
\end{figure}

From Fig.~\ref{fig:hits-rings-bebe}, the largest hit multiplicity is found in the two most inner BeBe rings. Thus, different BeBe rings configuration can be explored to optimize the centrality resolution of the proposed detector. In Fig.~\ref{fig:BBCentRingsConfiguration} different BeBe rings configurations were assumed to estimate the centrality resolution. For central collisions, the centrality resolution improves as the number of rings decreases from the inner to the outer. In this case, if we use the number of hits in the two outer rings is equivalent to the information given by the three inner rings (black and purple squares). For semi central and peripheral collisions, the BeBe centrality resolution is equivalent for all the six rings assumed configurations.  It is worth mentioning that even rings 4 and 5 give a good centrality resolution of 35\% for central collisions.

The centrality determination given by BeBe detector is fully complementary to the one that can be reached with the FHCAL ~\cite{Kurepin:2019bmh} detector at MPD-NICA, especially for central collisions where the FHCAL detector may lose resolution.

\subsection{Event plane resolution}
\label{sec:BeBeEventPlane}

\begin{figure}

\centering\begin{tikzpicture}[
            > = Straight Barb,
phasor/.style = { thick,-{Stealth}},
angles/.style = {draw, -, angle eccentricity=1,
                 right, angle radius=30mm}
                        ]
    \draw[->] (-0.5,0) -- (4,0) coordinate (x) node[below left] {$\mathit{p_{x}}$};
    \draw[->] (0,-0.5) -- (0,4) node[below left] (y) {$\mathit{p_{y}}$};
    \draw[phasor] (0,0) -- ( 60:4) coordinate (pT)  node[right] {$p_{T}$};
    \draw[phasor] (0,0) -- ( 30:4) coordinate (Q)  node[right] {$\vec{Q}$};
\coordinate (X)   at (0,0);
\draw
    pic["$\Psi_{RP}$",angles] {angle=x--X--Q}
    pic["$\varphi$'",angles] {angle=Q--X--pT}
    ;
\draw[ -] (1,0) arc (0:60:1);
\node (a) at (0.85,0.85) {$\varphi$};  

\end{tikzpicture}
\caption{Azimuth angle of particles in momentum coordinates $\varphi$, the reaction plane angle $\Phi_{RP}$ and $\varphi ' $ is the difference between the azimuth angle of particles and the reaction plane angle and $\vec{Q}$ is the vector used in the standard event plane angle method. }
\label{fig:anglesEP}
\end{figure}
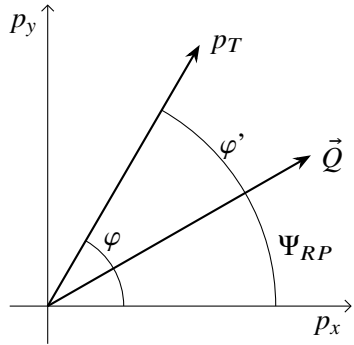

The BeBe detector aims to improve MPD's determination of the reaction plane, a key measurement for flow studies that provides physics insight into the early stages of the reaction. The information provided by BeBe can be used to study the anisotropic flow of particles produced in heavy-ion collisions which is typically quantified by the coefficients in the Fourier decomposition of the azimuthal angular particle distribution ~\cite{Voloshin:1994mz, Poskanzer:1998yz}. If the particle azimuthal angle is measured with respect to the direction of the reaction plane ~\cite{PhysRevC.77.034904}, then this Fourier analysis leads to
\begin{eqnarray}\label{eq:equation1}
 E \frac{dN}{d^3p} = \frac{1}{2\pi} 
 \frac{dN}{p_{\mathrm{T}}dp_{\mathrm{T}}d\eta} \left\lbrace 1 + 2 \sum_{n=1}^{\infty} v_{n} (p_{\mathrm{T}},\eta) \cos\left[n(\varphi - \Psi_n) \right]\right\rbrace, \nonumber \\
\end{eqnarray}
where $E$, $N$, $p$, $p_{\mathrm{T}}$, $\varphi$ and $\eta$ are the particle's energy, yield, total 3-momentum, transverse momentum, azimuthal angle and pseudo-rapidity, respectively. As shown in Fig.~\ref{fig:anglesEP} $\Psi_n$ is the reaction plane angle corresponding to the $n^{\mathrm{th}}$-order harmonic, $v_\mathrm{n}$. Experimentally, $\Psi_n$ can be determined using the sub-event correlation method discussed in Ref.~\cite{PhysRevC.58.1671}.

Profiting from the high granularity of the BeBe, we can resolve the event plane angle $\Psi^{BB}_n$ corresponding to the $n^{\mathrm{th}}$-order harmonic, using the reconstructed multiplicity provided by each disk cell of the hodoscope as follows~\cite{Voloshin:2008dg}
\begin{eqnarray} \label{eq:equation2}
\Psi^{BB}_n = \frac{1}{n} \, \tan^{-1} \left[ \sum_{i=1}^{m} w_i \, \sin (n\varphi_i){\mbox{\Huge{/}}} \sum_{i=1}^{m} w_i \, \cos (n\varphi_i)     \right], \nonumber \\
\label{EventPlane}
\end{eqnarray}
where $w_i$ is the multiplicity measured in the $i$-th cell, $m$ is the total number of BeBe cells and $\varphi_i$ is the $i$th-cell's azimuthal angle measured from the center of the hodoscope to the cell centroid.

\begin{figure}[!h]
\centering\includegraphics[width=0.7\linewidth]{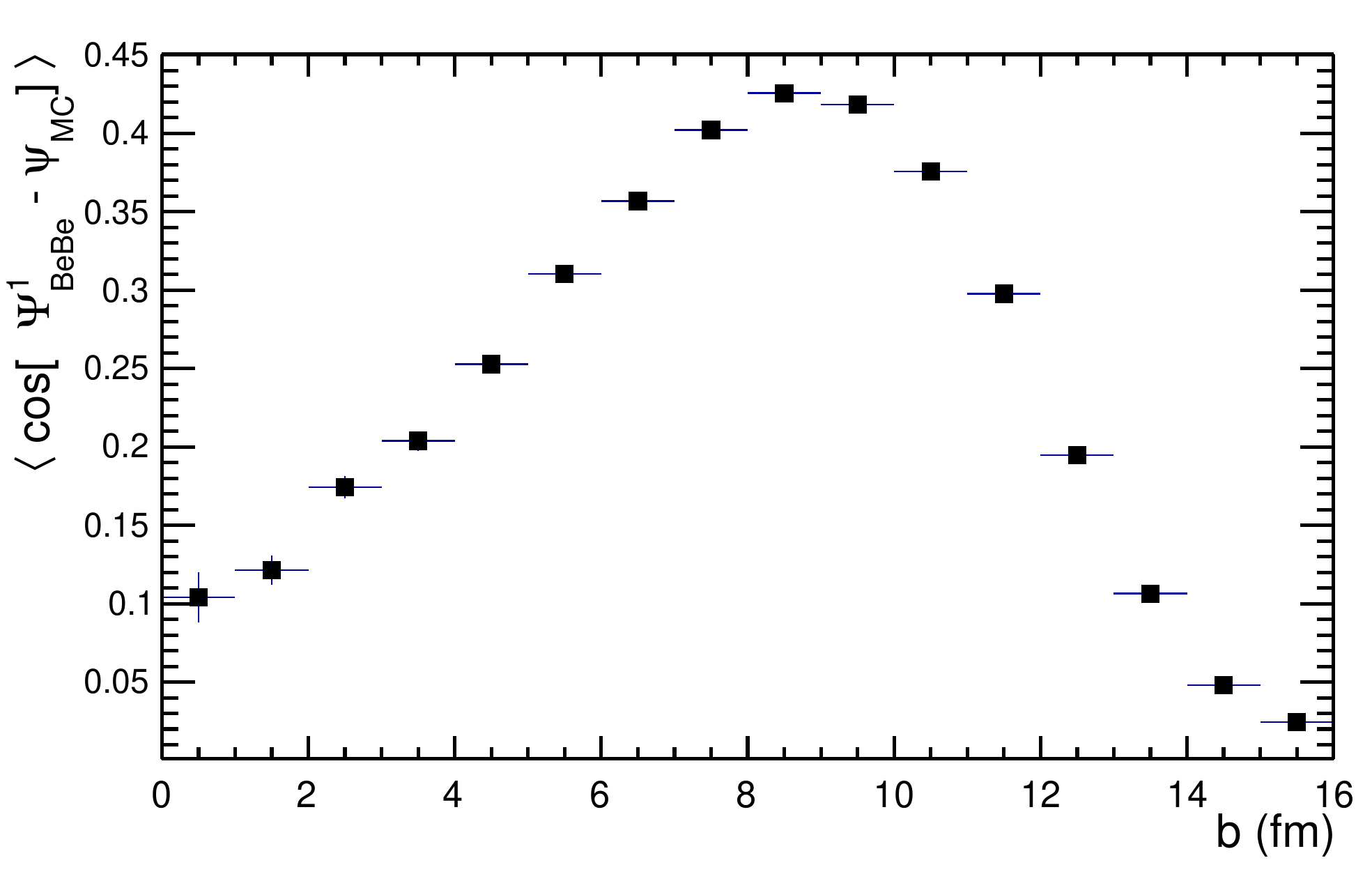}
\caption{Estimated event plane resolution using the BeBe detector.}
\label{fig:BBEventPlane}
\end{figure}

To estimate the event plane resolution with the proposed BeBe detector geometry, we simulated 1,000,000 minimum bias Bi+Bi collision events at $\sqrt{s_{NN}}=$ 9 GeV. The event generation was done with UrQMD, which includes multiple particle interactions, the excitation and fragmentation of color strings, and the formation and decay of hadron resonances, in the simulation of p+p, p+A, and A+A collisions. 
We used the MPD-ROOT offline framework~\cite{MPDROOT}. The produced particles were propagated through the detectors using GEANT-3 as a transport package. The multiplicity per cell, $w_i$, was estimated at hit-level and the event plane resolution with the BeBe detector for $n=1$ was computed as~\cite{Voloshin:2008dg}
\begin{equation}\label{eq:equation3}
\Big<\cos\Big(n\times (\Psi^{BB}_n-\Psi^{MC}_n)\Big)\Big >,
\end{equation}
where $\Psi^{MC}_n$ is the true value given by the Monte Carlo for the $n$-th order harmonic. Figure~\ref{fig:BBEventPlane} shows the dependence of the event plane resolution with the impact parameter for $n=1$. This effect has been also reported in Refs.~\cite{Abbas:2013taa, 1742-6596-742-1-012023,Ackermann:2000tr}. The BeBe is capable to reach a maximum of the event plane resolution for an impact parameter range of 6-11 for Bi+Bi collisions at $\sqrt{s_{NN}}=9$~GeV. 

\clearpage
\subsection{Conception of the BeBe readout electronics} %
\label{sec:BEBEreadout}

The simulation results show that the three outer rings of BeBe hodoscope give a good description of the centrality and event plane resolution. The number of signals generated (over a certain threshold in the time window previously described), can be used to obtain the number of charged particles ($N_{ch}$) that crosses those 48 cells in the rings 3, 4, and 5. For the trigger signal, all the rings were considered for the determination of the arrival time of BBL and BBR, denoted by TL and TE pulses, respectively.

\begin{figure}[hbt!]
\centering
\includegraphics[scale=0.4]{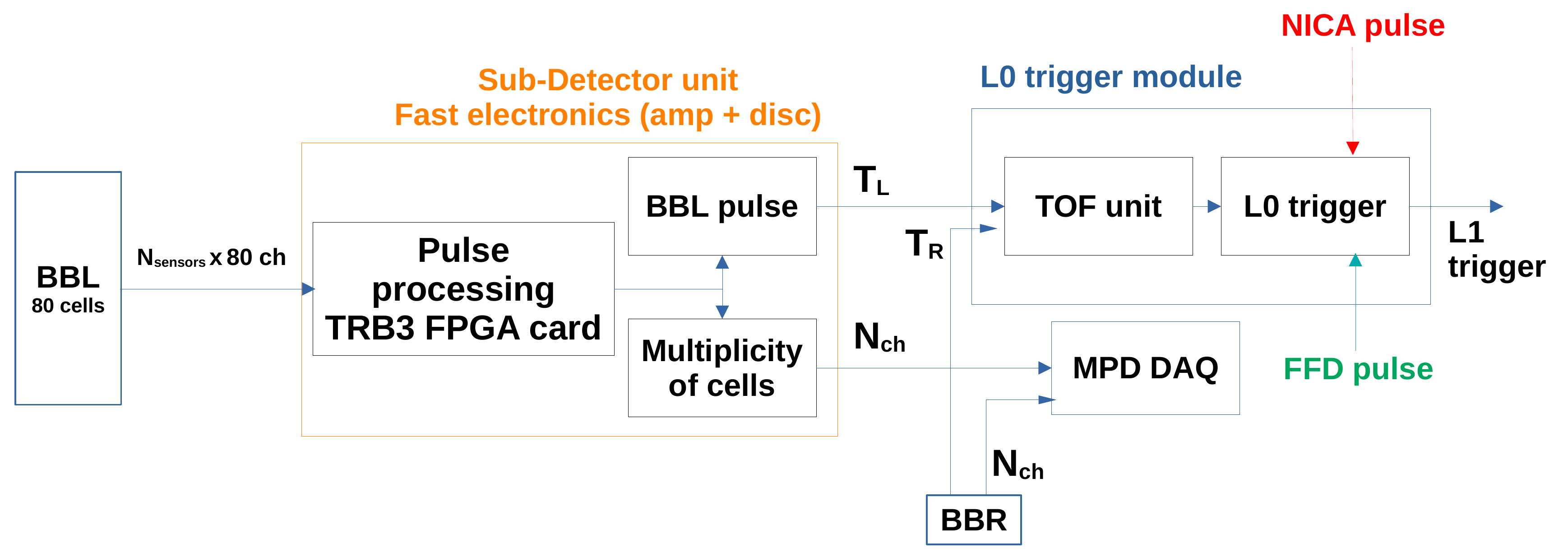} 
\caption{Schematic layout of the BeBe electronics.}
\label{fig:TriggerConcept}
\end{figure}

The signal collection for online and offline processing is schematically shown in Fig.\ref{fig:TriggerConcept}. All the trigger signals are collected by a TRB3 FPGA card, controlled by a Linux computer to acquire and store up to 264 input channels of information in a data center. Part of the signal processing task will be developed inside the FPGA card thus a single trigger for the Time Of Flight~(TOF) sensor will be generated inside this FPGA card, achieving the main objective of this front-end.

After processing the TL and TE pulses the L0 trigger pulse can be generated if the next two conditions are joined: 1. The time interval between the pulses TL and TR are into a selected window. 2. The NICA, FFD, and BeBe pulses are in coincidence.

\section{Conclusions}

The simulated geometry for BeBe detector proposal shows a good performance in triggering, event plane, and centrality determination. Our results suggest that at NICA energies the BeBe detector can be used for NICA beam monitoring in p+p and heavy-ion collisions with excellent trigger efficiencies for both systems. The maximum event plane resolution of BeBe is 43\% for an impact parameter range between 6 and 11 fm. For centrality determination, BeBe is a complementary detector to the FHCAL for central collisions. The BeBe detector could provide valuable information in heavy-ion collisions at NICA energies with the MPD.

The proposed BeBe detector is two plastic scintillator array stations located at $\pm$ 2 meters from the MPD interaction point. The plastic scintillator width is 1~cm. The proposed geometry of BeBe detector is similar to the one used in ALICE with the VZERO detector and its upgrade for the LHC Run 3, Fast Interaction Detector (FIT, V0+), a plastic scintillator disk segmented in 80 cells per station. 
For BeBe purposes, we wanted to develop this study in order to give a good trigger signal for the whole detector system (MPD). As a first approach, the time resolution of an individual BeBe cell ranges from 
0.47 and 1.39~ns depending on the number of photosensors attached to the cell. The development of the proposed data acquisition system for BeBe detector as described in section~\ref{sec:BEBEreadout} is a work in progress to be reported elsewhere. We hope this study can serve as a reference for future scopes of it. 

\acknowledgments

 The authors thank Prof. Alexey Kurepin for his valuable input and fruitful discussions about the potential of the BeBe detector proposal for MPD-NICA experiment. The authors are in debt to L. Díaz, E. Murrieta, and the offline group of MPD experiment for their technical support.  L.G.E.B. acknowledges support from postdoctoral fellowships granted by Consejo Nacional de Ciencia y Tecnolog\'ia. M.R.C. thankfully acknowledges computer resources, technical advice, and support provided by Laboratorio Nacional de Supercómputo del Sureste de México (LNS), a member of the CONACYT national laboratories, with project No. 53/2017. This work was partially supported by CONACyT research grants: A1-S-23238 and A1-S-13525.

\bibliographystyle{IEEEtran}
\bibliography{ref}


\end{document}